\documentclass[traditabstract]{aa}
\usepackage{graphicx}
\usepackage{natbib}
\usepackage{amsmath}
\usepackage{mwe,tikz}
\usepackage[percent]{overpic}

\usepackage[para,online,flushleft]{threeparttable}
\usepackage{txfonts}
\usepackage{hyperref}
\def\igr{IGR\,J17451--3022}
\def\inte{{\em INTEGRAL}}
\def\xmm{{\em XMM-Newton}}

\def\swift{{\em Swift}}

\def \inte {{\em INTEGRAL}}
\def \xmm {{\em XMM-Newton}}
\def \chandra {{$Chandra$}}
\def \suzaku {{$Suzaku$}}

\def \hcm {\hbox {\ifmmode $ atom cm$^{-2}\else atom cm$^{-2}$\fi}}

\def \arcsec {\hbox{$^{\prime\prime}$}}

\def \apjl {ApJL}
\def \apjs {ApJS}\def \aap {A\&A}

\begin{document}
   \title{IGR\,J17451--3022: a dipping and eclipsing low mass X-ray binary}

   \author{E. Bozzo
    \inst{1}
    \and P. Pjanka
    \inst{2}
   \and P. Romano
     \inst{3}        
   \and A. Papitto
     \inst{4}      
   \and C. Ferrigno
     \inst{1}   
   \and S. Motta 
     \inst{5}  
    \and A. A. Zdziarski 
      \inst{2}        
     \and F. Pintore
     \inst{6,7}  
      \and T. Di Salvo
     \inst{8}      
      \and L. Burderi
     \inst{6}          
     \and D. Lazzati
     \inst{9} 
     \and G. Ponti
     \inst{10} 
     \and L. Pavan  
     \inst{1} 
    }

   \institute{ISDC Data Centre for Astrophysics, Chemin d'Ecogia 16,
    CH-1290 Versoix, Switzerland; \email{enrico.bozzo@unige.ch}
    \and   
    Centrum Astronomiczne im.\ M. Kopernika, Bartycka 18, PL-00-716 Warszawa, Poland
    \and  
    INAF, Istituto di Astrofisica Spaziale e Fisica Cosmica - Palermo, via U. La Malfa 153, 90146 Palermo, Italy
    \and 
    Institut de Ci\`encies de l'Espai (IEEC-CSIC), Campus UAB, carrer de Can Magrans, S/N 08193, Cerdanyola del Vall\`es, Barcelona, Spain 
     \and 
    Astrohpysics, Department of Physics, University of Oxford, Keble Road, Oxford OX1 3RH, UK
    \and 
    Universit\'a degli Studi di Cagliari, Dipartimento di Fisica, SP Monserrato-Sestu, KM 0.7, 09042 Monserrato, Italy
    \and 
    INAF-Istituto di Astrofisica Spaziale e Fisica Cosmica - Milano, via E. Bassini 15, I-20133 Milano, Italy
    \and 
    Dipartimento di Fisica e Chimica, Universit\'a di Palermo, via Archirafi 36 - 90123 Palermo, Italy
    \and 
    Department of Physics, Oregon State University, 301 Weniger Hall, Corvallis, OR 97331, USA
    \and  
    Max-Planck-Institut f\"ur Extraterretrische Physik, Giessenbachstrasse, 85748 Garching, Germany
     }

   \date{}

  \abstract{In this paper, we report on the available X-ray data collected by \inte,\ \swift,\ and \xmm\ during the first outburst of the 
  \inte\ transient \igr,\ discovered in 2014 August. 
  The monitoring observations provided by the JEM-X instruments on-board \inte\ and the \swift\,/XRT showed that the event lasted 
  for about 9~months and that the emission of the source remained soft for the entire period. The source 
  emission is dominated by a thermal component ($kT\sim1.2~keV$), most likely produced by an accretion disk. The \xmm\ 
  observation carried out during the outburst revealed the presence of multiple absorption features in the soft X-ray emission that could be 
  associated to the presence of an ionized absorber lying above the accretion disk, as observed in many high-inclination low mass 
  X-ray binaries. The \xmm\ data also revealed the presence of partial and rectangular X-ray eclipses (lasting about 820~s), 
  together with dips. The latter can be associated with increases in the overall absorption column density in the direction of the source. 
  The detection of two consecutive  X-ray eclipses in the \xmm\ data allowed us to estimate the source orbital period at 
  $P_{\rm orb}$=22620.5$^{+2.0}_{-1.8}$~s (1$\sigma$ c.l.).}   
  

   \maketitle

\section{Introduction}
\label{sec:intro}

\igr\ is an X-ray transient discovered by \inte\ on 2014 August 22, during the observations performed in the direction of the 
Galactic center \citep{Chenevez14}. The source underwent a 9 months-long outburst and began its return to quiescence in the second 
half of May 2015 \citep{Bahramian15a}. At discovery, the source displayed a relatively soft emission, significantly detected by the 
JEM-X instrument on-board \inte\ only below 10\,keV \citep{Chenevez14}. The estimated 3--10 keV flux was $\sim$7~mCrab, corresponding to roughly 
10$^{-10}$~erg~cm$^{-2}$~s$^{-1}$. The outburst of \igr\ was followed-up with \swift\ \citep{Altamirano14, Heinke14, Bahramian15b, Bahramian15c, 
Bahramian15c}, \suzaku,\ \citep{Jaisawal15}, and \chandra\ \citep{Chakrabarty15}.  
The best determined source position was obtained from the \chandra\ data at 
RA(J$2000)$ = $17^{\rm h} 45^{\rm m} 06\fs72$, Dec(J$2000)=-30^{\circ} 22^{\prime} 43\farcs3$, 
with an associated uncertainty of 0\farcs6 (90\,\% c.l.). 
The spectral energy distribution of the source, together with the discovery of X-ray eclipses spaced by $\sim$6.3~h, led to the 
classification of this source as a low mass X-ray binary \citep[LMXB;][]{Bahramian15d, Jaisawal15}. 

In this paper we report on the analysis of all available \swift\ and \inte\ data collected during the outburst of the source, 
together with a \xmm\ Target of Opportunity observation (ToO) that was performed on 2014 March 6 for 40\,ks.  
\begin{figure}
  \includegraphics[width=8.5cm]{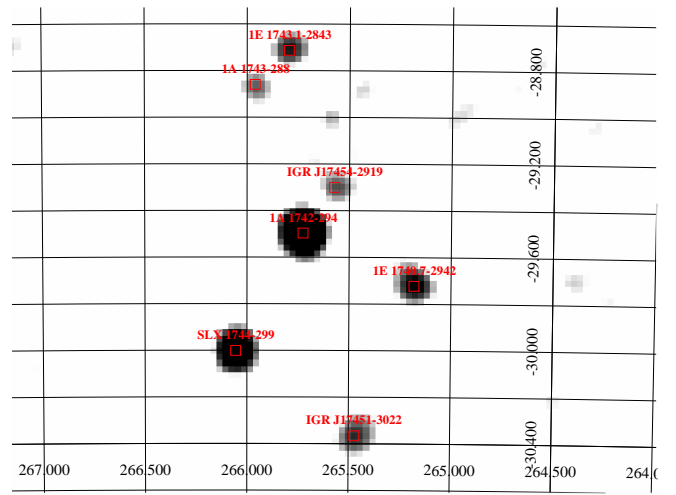}
  \caption{\inte\ JEM-X mosaicked image around the position of \igr\ (3-10~keV). This has been obtained by using the data 
  collected during the satellite revolutions 1448-1533 (excluding 1458, see text for details). The source is detected in 
  this mosaic with a significance of 23.7$\sigma$.}  
  \label{fig:mosa}
\end{figure}

\section{\inte\ data}
\label{sec:integral}

During its outburst, \igr\ was visible within the field of view of the two JEM-X units on-board 
\inte\ from satellite revolution 1448 (starting on 2014 August 18) to 1470 (hereafter first part of the outburst) 
and from 1508 to 1533 (ending on 2015 April 25; hereafter second part of the outburst). 
We analyzed all the publicly available \inte\ data and those for which our group got access rights in the 
\inte\ AO11 by using version 10.1 of the Off-line Scientific Analysis software   
(OSA) distributed by the ISDC \citep{courvoisier03}. \inte\ observations are divided into ``science windows'' (SCWs), 
i.e. pointings with typical durations of $\sim$2-3~ks. Only SCWs in which 
the source was located to within an off-axis angle of 4.5~deg from the center of the JEM-X field of view 
\citep{lund03} were included in the JEM-X analysis, 
while for IBIS/ISGRI we retained all SCWs where the source was within an off-axis angle of 12~deg from the center 
of the instrument field of view.  
We extracted the IBIS/ISGRI \citep{ubertini03,lebrun03} mosaics in the 20-40~keV and 40-80~keV energy bands, while  
the JEM-X mosaics were extracted in the 3-10~keV and 10-20~keV energy bands (we used below JEM-X1 as a reference to provide 
all quantitative results but checked that compatible values, to within the uncertainties, would be obtained with JEM-X2).  

By checking the IBIS/ISGRI mosaics of each revolution, we noticed that the source was never significantly 
detected except for revolution 1458 which covers the period 2014 September 21 at 19:02 UTC to September 24 at 
03:24 UTC (total exposure time 158~ks). In this revolution, we estimated from the IBIS/ISGRI mosaics 
a source detection significance of 8.4$\sigma$ in the 20-40~keV energy band and 7.8$\sigma$ in the 40-80~keV energy band. 
The corresponding source count-rate in the two energy bands was  
0.80$\pm$0.10~cts~s$^{-1}$ and 0.54$\pm$0.07~cts~s$^{-1}$, which translate into fluxes of 5.7$\pm$0.7~mCrab 
($\sim$4.4$\times$10$^{-11}$~erg~cm$^{-2}$~s$^{-1}$) and 7.2$\pm$0.9~mCrab 
($\sim$5.0$\times$10$^{-11}$~erg~cm$^{-2}$~s$^{-1}$), respectively\footnote{The conversion from count-rate to mCrab 
was carried out by using the most recent observations of the Crab (at the time of writing) in spacecraft revolution 1597. 
From these data we measured for the Crab a count-rate of 139.4~cts~s$^{-1}$ and 75.2~cts~s$^{-1}$ from the IBIS/ISGRI mosaics in the 
20-40~keV and 40-80~keV. The estimated count-rates in the JEM-X mosaics were of 111.6~cts~s$^{-1}$ and 29.6~cts~s$^{-1}$ in the energy band 
3-10~keV and 10-20~keV, respectively.}. 
In the IBIS/ISGRI mosaic obtained during the first part of the outburst (revolutions 1448-1470), 
excluding revolution 1458, the source was only marginally detected at $\sim$5.6$\sigma$ in the 20-40~keV energy range and 
$\sim$7.0$\sigma$ in the 40-80~keV energy range. The count-rate of the source was roughly 0.2~cts~s$^{-1}$ in both energy bands, 
corresponding to a flux of $\sim$1.5~mCrab in the 20-40~keV energy band and $\sim$2.5~mCrab in the 40-80~keV energy band 
(i.e. $\sim$1.2$\times$10$^{-11}$~erg~cm$^{-2}$~s$^{-1}$ and $\sim$1.7$\times$10$^{-11}$~erg~cm$^{-2}$~s$^{-1}$, respectively).  
Given the relatively long cumulated exposure time of the mosaic (1.4 Ms), several structures were detected at significances 
close or higher than that of  \igr.\ 
For this reason, we cannot discriminate if the fluxes measured from the position of the source are genuine or simply due to 
background fluctuations arising in the crowded and complex region of the Galactic Center. 
We can certainly conclude that the overall flux of the source during the first part of the outburst in the IBIS/ISGRI energy band 
was significantly lower than that measured during the revolution 1458 alone. 
A similar conclusion applies to the IBIS/ISGRI mosaic 
extracted during the second part of the outburst, i.e. including data from revolution 1508 to 1533 (total exposure time 1.3~Ms). 
In this case the detection significance of the source was of only 5.0$\sigma$ in the 20-40 keV energy band and 
3.6$\sigma$ in the 40-80~keV energy range. These low significance detections are likely due to background fluctuations. 
We estimated in this case a 3$\sigma$ upper limit on the source flux of about 2~mCrab in both the 20-40~keV and 40-80~keV energy bands 
(corresponding to $\sim$1.5$\times$10$^{-11}$~erg~cm$^{-2}$~s$^{-1}$ and $\sim$1.4$\times$10$^{-11}$~erg~cm$^{-2}$~s$^{-1}$, respectively). 
By combining the ISGRI data from all revolution 1448-1533, excluding revolution 1458, we did not notice an increase in the source 
detection significance. In this deeper mosaic (total exposure time 2.7~Ms), the source detection 
was at 6$\sigma$ in the 20-40~keV energy band and 7.6$\sigma$ in the 40-80~keV energy band. This finding 
supports the idea that the mentioned low level detections in all revolutions but 1458 are likely to 
result from background fluctuations. 

In revolution 1458, the source was detected at 5.8$\sigma$ 
in the 3-10~keV JEM-X energy band with a count-rate of 0.80$\pm$0.14~cts~s$^{-1}$ (total exposure time 37~ks). 
The estimated flux was 7.2$\pm$1.2~mCrab ($\sim$1.0$\times$10$^{-10}$~erg~cm$^{-2}$~s$^{-1}$). 
The source was only marginally detected ($\sim$3$\sigma$) in the 10-20~keV energy band with a count-rate of 
0.17$\pm$0.06~cts~s$^{-1}$ and a flux of 5$\pm$2~mCrab (corresponding to 
$\sim$4.6$\times$10$^{-11}$~erg~cm$^{-2}$~s$^{-1}$). 

We did not detect significant variability of the 3-10~keV source flux in the JEM-X data collected 
in all other revolutions, but the corresponding flux in the 10-20~keV energy band seemed to decrease 
by a factor of at least $\gtrsim$3. In the JEM-X mosaic 
built by using data from revolution 1448 to 1470 (excluding revolution 1458), the source was detected with a count-rate of 
0.67$\pm$0.04~cts~s$^{-1}$ (detection significance 16$\sigma$) in the 3-10~keV energy band and we estimated a 
3$\sigma$ upper limit of 2~mCrab for the flux in the 10-20~keV energy band (total exposure time 367~ks). 
In revolutions 1508 to 1533, the source was detected with a count-rate of 
0.70$\pm$0.04~cts~s$^{-1}$ (detection significance 18$\sigma$) in the 3-10~keV energy band and we estimated a 
3$\sigma$ upper limit of 2~mCrab for the flux in the 10-20~keV energy band (total exposure time 372~ks). 
When data from all revolutions but 1458 are merged (see Fig.~\ref{fig:mosa}), the estimated 
source count-rate in the 3-10~keV energy range is 0.68$\pm$0.03~cts~s$^{-1}$ (detection significance 23.7$\sigma$) 
and the 3$\sigma$ upper limit for the flux in the 10-20~keV energy band is 1~mCrab (total exposure time 738~ks).  
 
\begin{figure}
  \includegraphics[width=6.1cm, angle=-90]{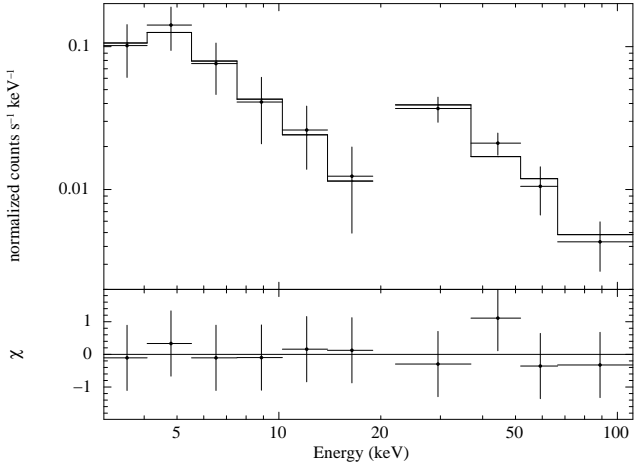}
  \caption{Combined JEM-X1 and ISGRI spectrum of \igr\ extracted from the \inte\ data in revolution 1458. 
  The best fit model is obtained by using an absorbed power-law model (see text for details). 
  The residuals from the fit are shown in the bottom panel.}  
  \label{fig:jmxspectra}
\end{figure}

Given the above results, we focused on the JEM-X and ISGRI spectra extracted from the data collected during satellite revolution 
1458. As the source was relatively faint and the surrounding region crowded, the JEM-X spectra were extracted with the {\sc mosaic\_spec} 
tool from a mosaic with 8 energy bins (spanning the range 3-25 keV) in order to avoid contamination from nearby sources. 
ISGRI spectra were extracted by using a 6 energy bins response matrix in the range 20-100~keV. We ignored during the fit to these spectra 
the energy range 20-25~keV in JEM-X due to the lack of signal from the source and the first bin in ISGRI to avoid calibration uncertainties. 
The combined JEM-X+ISGRI spectrum could be well fit ($\chi^2_{\rm red}$/d.o.f.=0.4/7) 
by using a simple absorption power-law model (the absorption column density was fixed to the best value 
determined from the \xmm\ data, i.e. $N_{\rm H}$=5.6$\times$10$^{22}$~cm$^{-2}$; see Sect.~\ref{sec:xmm}). 
The best fit power-law photon index was $\Gamma$=2.0$\pm$0.4 and the measured absorbed flux in the 3-20~keV (20-100~keV) was  
1.0$\times$10$^{-10}$~erg~cm$^{-2}$~s$^{-1}$ (8.8$\times$10$^{-11}$~erg~cm$^{-2}$~s$^{-1}$). A fit with a single thermal 
component, as {\sc bbodyrad} or {\sc diskbb} in {\sc Xspec}, could not provide an acceptable fit to the broad-band 
data ($\chi^2_{\rm red}$/d.o.f.$\gtrsim$7). The combined JEM-X+ISGRI spectrum of \igr\ is shown in Fig.~\ref{fig:jmxspectra}.

Due to the low signal-to-noise ratio, no meaningful ISGRI spectrum could be extracted for the source when all data but those 
in revolution 1458 were stacked together. The corresponding JEM-X spectrum resulted in 
only 3 energy bins with a significant flux measurement and thus discriminating between different spectral models (i.e. a power-law rather than 
a thermal component) was not possible. The spectrum would be compatible with both a power-law of photon index $\sim$3 or a {\sc diskbb} with 
a temperature of $\sim$1~keV. The 3-10~keV flux estimated from the spectral fit was $\sim$1.0$\times$10$^{-10}$~erg~cm$^{-2}$~s$^{-1}$ 
(not corrected for absorption). 

Considering all the results of the \inte\ analysis together, we thus conclude that the spectral energy distribution measured during the 
first detectable outburst of \igr\ remained generally soft and hardly detected above $\sim$10~keV. There are, however, indications that the 
source might have undergone a spectral state transition during the \inte\ revolution 1458, where a significant detection above 20~keV 
was recorded. 

Finally, we also extracted the JEM-X1 and JEM-X2 lightcurves with a time resolution of 2~s to search for type-I X-ray bursts. 
However, we did not find any significant detection.

\section{\swift\ data}
\label{sec:swift}

\swift\,/XRT \citep{burrows05} periodically observed the source starting close to its 
initial detection, until it faded into quiescience 9 months later (i.e. from 2014 August 5 to 
2015 June 6). Observations in the period 2014 November to 2015 early February were not 
available due to the satellite Sun constraints.  
A log of all available XRT observations is reported in Table~\ref{tab:swift}. 

XRT data were collected in both windowed-timing (WT) and photon-counting (PC) 
mode and analyzed by using standard procedures and the FTOOLS software (v6.16).   
The XRT data were processed with the {\sc xrtpipeline} 
(v.0.13.1) and the latest calibration files available (20140730). 
The WT data were never affected by pile-up, therefore 
WT source events were extracted from circular regions of 20 pixels 
(1 pixel $\sim2.36$\arcsec). On the contrary, when required, 
the PC data were corrected for pile-up by adopting standard procedures 
\citep{vaughan2006:050315,Romano2006:060124}. In particular, we determined  
the size of the affected core of the point spread function (PSF) by comparing 
the observed and nominal PSF, and excluded from the analysis all the events that fell 
within that region. The PC source events were thus extracted either from an annuli with 
an outer radius of 20 pixels and an inner radius determined from the  
severity of the pile-up, or within circular regions with a radius of 20 pixels when pile-up was not 
present.  Background events were extracted from nearby source-free regions. 
%Source and background event lists were barycentered by using the {\sc barycorr} tool. 
From each XRT observation we extracted an averaged spectrum and the corresponding barycentered lightcurve.  
The time resolution of the lightcurves was 1~s for data in WT mode and 2.5~s for data in PC mode. The 
barycentric correction was applied by using the {\sc barycorr} tool.  
For the observations 025--028, a single spectrum was extracted due to the low statistics of these data. 
All spectra were binned to ensure at least 20 counts per energy bin 
and were fit in the 0.5--10\,keV energy range. 

All XRT spectra could be well fit by using a simple absorbed disk blackbody model 
({\sc tbabs}*{\sc diskbb} in {\sc Xspec}; see also Sect.~\ref{sec:xmm}). 
Fits with a power-law model provided in most cases significantly worst results. 
We report the best fit obtained in all cases in Table~\ref{tab:swift}.  
The XRT monitoring showed that the spectral energy distribution of the 
X-ray emission from \igr\ was roughly stable along the entire outburst. 
The absorption column density and the temperature of the thermal component 
measured by XRT remained always compatible (to within the uncertainties) 
with the values obtained from the fit to the \xmm\ data (see \ref{sec:xmm}). We verified that fixing the XRT absorption 
column density to the \xmm\ value would result in acceptable fits to all observations and negligible changes in the 
temperature of the thermal component. We note that no XRT observations 
were performed during (or sufficiently close in time to) the \inte\ revolution 1458, 
where a possible spectral state transition of the source was observed (see Sect.~\ref{sec:integral}).  
 \begin{table*} 	
 \begin{center} 
 \scriptsize
 \caption{Log of all \swift\/XRT observations use in this work. For each observation we also report the results of the spectral fit 
 obtained by using an absorbed {\sc diskbb} model (see Sect.~\ref{sec:swift} and \ref{sec:xmm}). All uncertainties are given at 90\% c.l.}   	
  \label{tab:swift} 	
 \begin{tabular}{llllllllll} 
 \hline 
 \hline 
 \noalign{\smallskip} 
 Sequence   & Obs/Mode  & Start time  (UT)  & End time   (UT) & Exposure & $N_{\rm H}$ & $kT_{\rm diskbb}$ & $N_{\rm diskbb}$ & $F_{\rm 0.5--10\,keV}$$^a$ & $\chi_{\rm red}^2$/d.o.f. \\ 
   &     & (UTC)  & (UTC)  & (s) & (10$^{22}$~cm$^{-2}$) & (keV)  &   &  (10$^{-10}$~erg~cm$^{-2}$~s$^{-1}$) &    \\
 \noalign{\smallskip} 
 \hline 
 \noalign{\smallskip} 
00037734002	&	XRT/PC	&	2014-09-05 15:28:19	&	2014-09-05 18:48:54	&	1655	& 4.6$\pm$0.6 & 1.0$\pm$0.1 & 11.6$^{+10.1}_{-5.2}$ & 0.79$^{+0.06}_{-0.06}$ & 1.16/35 	\\
00037734003	&	XRT/PC	&	2014-09-09 10:03:10	&	2014-09-09 10:35:53	&	1963	& 5.7$^{+0.4}_{-0.6}$ & 1.2$\pm$0.1 & 8.9$^{+6.8}_{-3.8}$ & 1.06$^{+0.07}_{-0.07}$ & 0.86/51 	\\
00037734005	&	XRT/PC	&	2014-09-10 02:15:12	&	2014-09-10 12:09:56	&	1730	& 5.7$^{+0.9}_{-0.8}$ & 1.1$\pm$0.1 & 9.0$^{+9.7}_{-4.6}$ & 0.69$^{+0.06}_{-0.06}$ & 0.83/31 	\\
00037734006	&	XRT/WT	&	2014-09-10 18:32:26	&	2014-09-10 23:03:06	&	1980	& 5.2$\pm$0.5 & 1.4$\pm$0.1 & 3.6$^{+2.0}_{-1.2}$ & 0.90$^{+0.05}_{-0.05}$ & 0.79/94 	\\
00037734007	&	XRT/WT	&	2014-09-11 16:50:33	&	2014-09-11 18:43:31	&	1992	& 5.4$\pm$0.4 & 1.11$\pm$0.06 & 13.0$^{+4.7}_{-3.4}$ & 1.20$^{+0.04}_{-0.04}$ & 1.03/147 	\\
00037734008	&	XRT/PC	&	2014-09-15 06:45:57	&	2014-09-15 08:25:08	&	955	& 6.0$\pm$1.0 & 1.2$\pm$0.2 & 11.5$^{+13.0}_{-5.9}$  & 1.44$^{+0.13}_{-0.13}$ & 0.80/29 	\\
00037734010	&	XRT/PC	&	2014-09-19 15:05:26	&	2014-09-19 15:22:54	&	1028	& 6.7$^{+1.6}_{-1.3}$ & 1.0$\pm$0.2 & 42.4$^{+77.2}_{-26.5}$ & 1.50$^{+0.14}_{-0.15}$ & 0.80/18 	\\
00037734013	&	XRT/PC	&	2014-09-25 00:28:07	&	2014-09-25 00:44:55	&	1008	& 9.2$\pm$4.0 & 2.0$^{+1.0}_{-0.6}$ & 0.3$^{+1.3}_{-0.3}$ & 0.33$^{+0.05}_{-0.06}$ & 0.44/13 \\
00033439001	&	XRT/WT	&	2014-09-25 14:48:06	&	2014-09-25 15:00:13	&	724	& 4.7$\pm$0.7 & 1.5$\pm$0.2 & 2.5$^{+2.2}_{-1.2}$ & 1.18$^{+0.09}_{-0.09}$ & 1.18/51 	\\
00033439003	&	XRT/PC	&	2014-09-29 09:49:40	&	2014-09-29 10:06:56	&	1020	& 6.5$\pm$1.0 & 1.3$\pm$0.2 & 9.7$^{+9.6}_{-4.8}$ & 1.63$^{+0.12}_{-0.13}$ & 0.98/38 	\\
00033439004	&	XRT/PC	&	2014-10-03 05:08:51	&	2014-10-03 10:03:54	&	1111	& 6.4$\pm$1.3 & 1.4$\pm$0.2 & 4.6$^{+5.9}_{-2.6}$ & 1.36$^{+0.12}_{-0.13}$ & 0.46/25 	\\
00033439007	&	XRT/WT	&	2014-10-09 01:55:37	&	2014-10-09 02:10:57	&	904	& 7.4$^{+4.7}_{-2.7}$ & 1.7$^{+1.0}_{-0.5}$ & 0.5$^{+3.6}_{-0.5}$ & 0.36$^{+0.07}_{-0.08}$ & 0.61/18 \\
00033439008	&	XRT/WT	&	2014-10-12 08:06:00	&	2014-10-12 08:25:58	&	1172	& 6.0$^{+1.0}_{-0.8}$ & 1.4$\pm$0.2 & 8.4$^{+8.0}_{-4.0}$ & 2.05$^{+0.16}_{-0.17}$ & 0.93/54 	\\
00033439009	&	XRT/WT	&	2014-10-15 00:07:46	&	2014-10-15 00:23:29	&	936	& 6.3$\pm$0.6 & 1.2$\pm$0.1 & 15.6$^{+8.7}_{-5.5}$ & 1.82$^{+0.07}_{-0.08}$ & 0.98/100 	\\
00033439010	&	XRT/WT	&	2014-10-18 04:47:25	&	2014-10-18 05:03:58	&	971	& 5.9$\pm$0.6 & 1.2$\pm$0.1 & 12.3$^{+6.7}_{-4.3}$ & 1.57$^{+0.07}_{-0.08}$ & 1.11/87 	\\
00033439011	&	XRT/WT	&	2014-10-21 13:03:02	&	2014-10-21 13:19:58	&	1005	& 6.1$\pm$0.6 & 1.09$\pm$0.07 & 22.6$^{+11.0}_{-7.2}$ & 1.70$^{+0.08}_{-0.08}$ & 1.09/104 	\\
00033439012	&	XRT/WT	&	2014-10-24 11:02:12	&	2014-10-24 11:18:57	&	999	& 5.9$\pm$0.6 & 1.1$\pm$0.1 & 17.4$^{+11.9}_{-6.9}$ & 1.36$^{+0.06}_{-0.06}$ & 0.93/89 	\\
00033439014	&	XRT/PC	&	2015-02-03 05:16:37	&	2015-02-03 06:51:54	&	1991	& 5.4$\pm$0.6 & 1.4$\pm$0.2 & 5.3$^{+3.6}_{-2.2}$ & 1.78$^{+0.12}_{-0.12}$ & 1.31/56 	\\
00033439015	&	XRT/WT	&	2015-02-17 19:00:34	&	2015-02-17 19:18:58	&	1100	& 7.1$\pm$0.9 & 1.4$\pm$0.1 & 6.4$^{+4.5}_{-2.6}$ & 1.42$^{+0.08}_{-0.09}$ & 1.49/84 	\\
00033439016	&	XRT/WT	&	2015-03-03 21:34:47	&	2015-03-03 21:52:57	&	1079	& 5.8$\pm$0.6 & 1.3$\pm$0.1 & 9.7$^{+4.4}_{-3.0}$ & 1.84$^{+0.08}_{-0.08}$ & 1.11/113 	\\
00033439017	&	XRT/WT	&	2015-03-17 13:01:51	&	2015-03-17 13:13:58	&	706	& 6.0$\pm$0.7 & 1.0$\pm$0.1 & 23.4$^{+17.8}_{-9.8}$ & 1.31$^{+0.08}_{-0.08}$ & 1.2/61 	\\
00033439018	&	XRT/WT	&	2015-03-30 14:02:44	&	2015-03-30 14:17:57	&	901	& 4.7$\pm$0.5 & 1.1$\pm$0.1 & 12.4$^{+6.9}_{-4.4}$ & 1.34$^{+0.07}_{-0.07}$ & 1.08/82 	\\
00033439020	&	XRT/PC	&	2015-04-24 00:03:17	&	2015-04-24 03:26:56	&	1504	& 5.3$\pm$1.4 & 1.1$\pm$0.2 & 6.7$^{+10.5}_{-4.1}$ & 0.53$^{+0.05}_{-0.06}$ & 0.90/20 	\\
00033439021	&	XRT/PC	&	2015-04-24 04:36:39	&	2015-04-24 05:04:54	&	1675	& 5.4$^{+1.7}_{-1.3}$ & 1.6$^{+1.0}_{-0.5}$ & 0.3$^{+1.7}_{-0.3}$ & 0.19$^{+0.03}_{-0.05}$ & 1.21/13 \\
00033439022	&	XRT/PC	&	2015-04-24 08:03:36	&	2015-04-24 08:24:54	&	1269	& 4.8$\pm$1.0 & 1.1$\pm$0.1 & 8.6$^{+7.7}_{-4.0}$ & 0.81$^{+0.05}_{-0.06}$ & 1.01/35	\\
00033439023	&	XRT/WT	&	2015-05-09 08:53:36	&	2015-05-09 09:07:57	&	839	& 5.1$\pm$1.0 & 0.86$\pm$0.1 & 20.7$^{+27.8}_{-11.3}$ & 0.48$^{+0.04}_{-0.04}$ & 1.22/33 	\\
00033439025$^{b}$	&	XRT/PC	&	2015-05-18 22:35:51	&	2015-05-18 23:09:04	&	1986  & 3.7$\pm$0.6 & 1.0$\pm$0.1 & 0.98$^{+1.1}_{-0.5}$ & 0.046$^{+0.004}_{-0.004}$ & 1.22/244	\\
00033439026$^{b}$	&	XRT/PC	&	2015-05-23 11:10:38	&	2015-05-23 13:16:56	&	697	& --- & --- & --- & --- & --- 	\\
00033439027$^{b}$	&	XRT/PC	&	2015-05-30 09:24:15	&	2015-05-30 11:00:54	&	980	& --- & --- & --- & --- & --- 	\\
00033439028$^{b}$	&	XRT/PC	&	2015-06-06 10:53:18	&	2015-06-06 18:58:55	&	1667	& --- & --- & --- & --- & --- 	\\
\hline
\hline
  \end{tabular}
  \end{center}
  \begin{list}{}{} 
  \scriptsize
  \item[$^{\mathrm{a}}$:] Observed flux in units of 10$^{-10}$~erg~cm$^{2}$~s$^{-1}$ (not corrected for absorption) 
   \item[$^{\mathrm{b}}$:] We merged data from observations 00033439025 to 00033439028, as in all these cases the statistics was 
   too low to carry out separated meaningful spectral fits. The results obtained from the fits to the merged observations is reported on the row 
   corresponding to the observation ID.~00033439025.  
  \end{list} 
  \end{table*}

\begin{figure}
  \includegraphics[width=6.5cm, angle=-90]{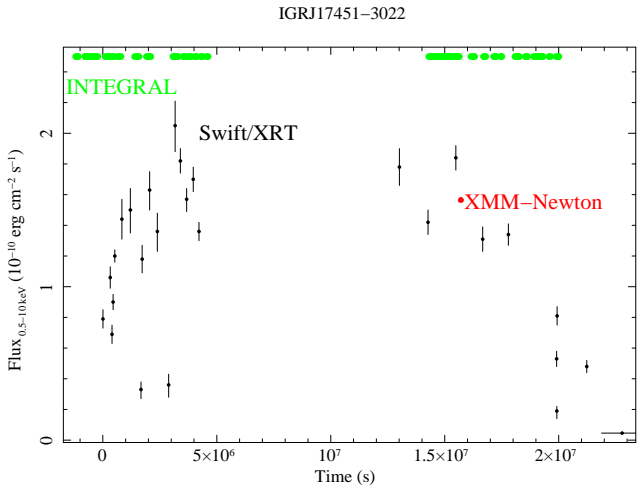}
  \caption{\swift\,/XRT lightcurve of the entire outburst recorded from \igr.\ The flux is not corrected for absorption. 
  The start time of the lightcurve (t=0) is 2014 September 
  5 at 15:37 (UTC). We also show in the plot the coverage provided by \inte\ (see Sect.~\ref{sec:integral}) and 
  the flux measured by \xmm\ (see Sect.~\ref{sec:xmm}).}  
  \label{fig:swift}
\end{figure}

Given the relatively low statistics of each XRT observation, we did not perform a detailed timing analysis of the data. 
However, we show in Fig.~\ref{fig:swift} the total outburst lightcurve of the source as observed by XRT from the onset of the event 
down to the return in quiescence. We marked on this figure the time intervals corresponding to the coverage provided by \inte\ and 
the time of the ToO observation carried out by \xmm\ (see Sect.~\ref{sec:xmm}). 

\section{\xmm\ data}
\label{sec:xmm}

The \xmm\ \citep{jansen01} ToO observation of \igr\ was performed on 2015 March 06 for a total exposure time of 40~ks. 
The EPIC-pn was operated in timing mode, the MOS2 in full frame, and the MOS1 in small window. 
We reduced these data by using the SAS version 14.0 and the latest \xmm\ calibrations files 
available\footnote{\href{http://xmm2.esac.esa.int/external/xmm_sw_cal/calib/index.shtml}
{http://xmm2.esac.esa.int/external/ xmm\_sw\_cal/calib/index.shtml}} \citep[XMM-CCF-REL-329; see also][]{pintore14}. 
The observation was slightly affected by intervals of 
high flaring background and thus we filtered out these intervals following the SAS online 
thread\footnote{\href{http://xmm.esac.esa.int/sas/current/documentation/threads/}{http://xmm.esac.esa.int/sas/current/documentation/threads/}}. 
The final effective exposure time for the three EPIC cameras was 29~ks. 

\begin{figure}
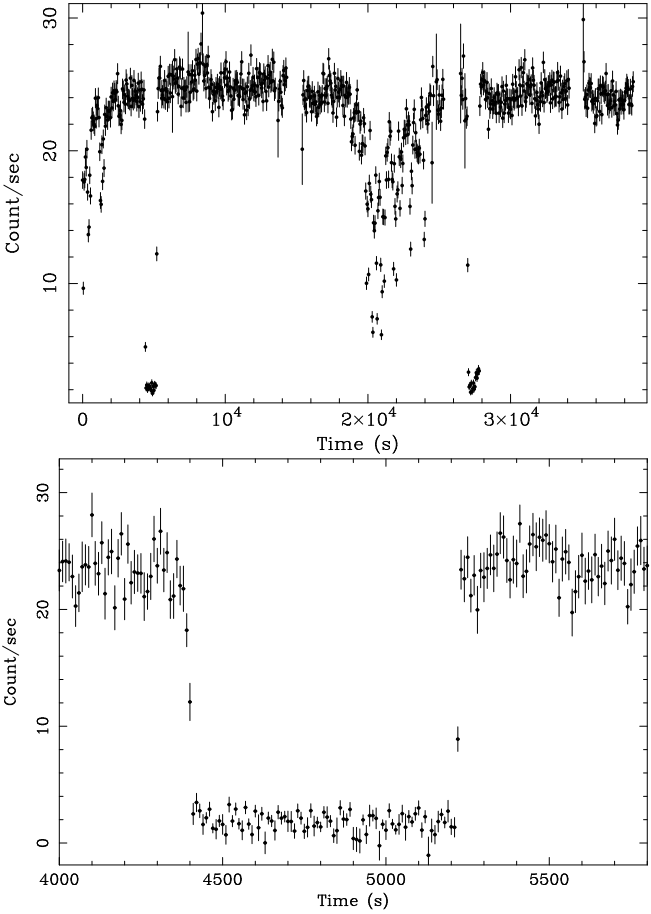

 \centering
     \includegraphics[width=6cm, angle=-90]{lcurve_xmm.ps}
     \includegraphics[width=6cm,angle=-90]{eclipses.ps}
  \caption{{\it Upper figure}: EPIC-pn lightcurve of \igr\ (0.5-12~keV). The start time of the lightcurve (t=0) is 2015 March 6 at  
  6:26 (UTC). The bin time is 50~s for display purposes. Time intervals with no data are due to the filtering of the high 
  flaring background. {\it Bottom figure}: a zoom of the first eclipse (time bin 10~s). The apparent different count-rate 
  at the bottom of the eclipse between this and the upper figure is due to the different time binning.}
  \vspace{-0.5cm}
    \label{fig:lcurve_xmm}
\end{figure}

\begin{figure}
 \centering
 \vspace{0.5cm}
    \includegraphics[width=6.0cm,angle=-90]{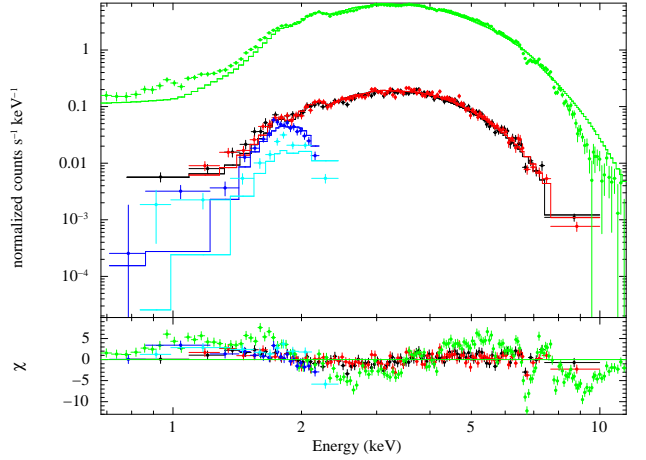}
  \caption{EPIC-pn (green), EPIC-MOS1 (black), EPIC-MOS2 (red), RGS1 (blue), and RGS2 (cyan) average spectra 
  extracted from the \xmm\ observation of \igr.\ The model used to preliminary fit the data here is an 
  absorbed disk blackbody model ({\sc tbabs*diskbb} in {\sc Xspec}). The residuals from the fit in the bottom 
  panel show the instrumental excess in the EPIC-pn data below 1.7~keV (beside the many absorption features, see 
  also Fig.~\ref{fig:ratio}). For this reason, we only 
  used EPIC-pn data in the energy range 1.7-12~keV for further analysis.} 
    \label{fig:residuals}
\end{figure}

We first extracted the EPIC-pn lightcurve of the source in the energy range 0.5-12~keV from CCD columns 28-46. 
The corresponding background lightcurve was extracted by using columns 3-10. The background-subtracted  
EPIC-pn lightcurve of \igr\ with a time bin of 50~s is shown in Fig.~\ref{fig:lcurve_xmm} 
(we also used the {\sc epiclccorr} tool to correct the lightcurves for any relevant instrumental effect). The source displayed 
an average count-rate of 16~cts~s$^{-1}$, two X-ray eclipses and the presence of dips at the beginning 
of the observation and between the two eclipses. Given the source count-rate, both MOS1 and MOS2 data suffered from significant 
pile-up. This was corrected by following the SAS online threads. The RGS spectra were characterized by a low statistics, due to the 
significant extinction recorded in the direction of the source (see later in this section). We thus used the RGS and the MOS data 
only to check consistency with the results obtained from the analysis of the EPIC-pn data, but do not discuss them in details. 
A fit of all data together revealed the presence of instrumental residuals in the EPIC-pn below 1.7~keV (see Fig.~\ref{fig:residuals}). 
This issue has been reported in several papers using the EPIC-pn in timing mode and thus we restricted our spectral analysis to the 
1.7-12~keV energy range \citep[see, e.g.,][for a similar issue with EPIC-pn residuals below 1.7~keV]{piraino12,pintore15,dai15,boirin05}. 
We barycentered the EPIC-pn data before carrying out any timing analysis. 

\subsection{The X-ray eclipses}
\label{sec:eclipses}

\begin{figure}
  \includegraphics[width=8.5cm]{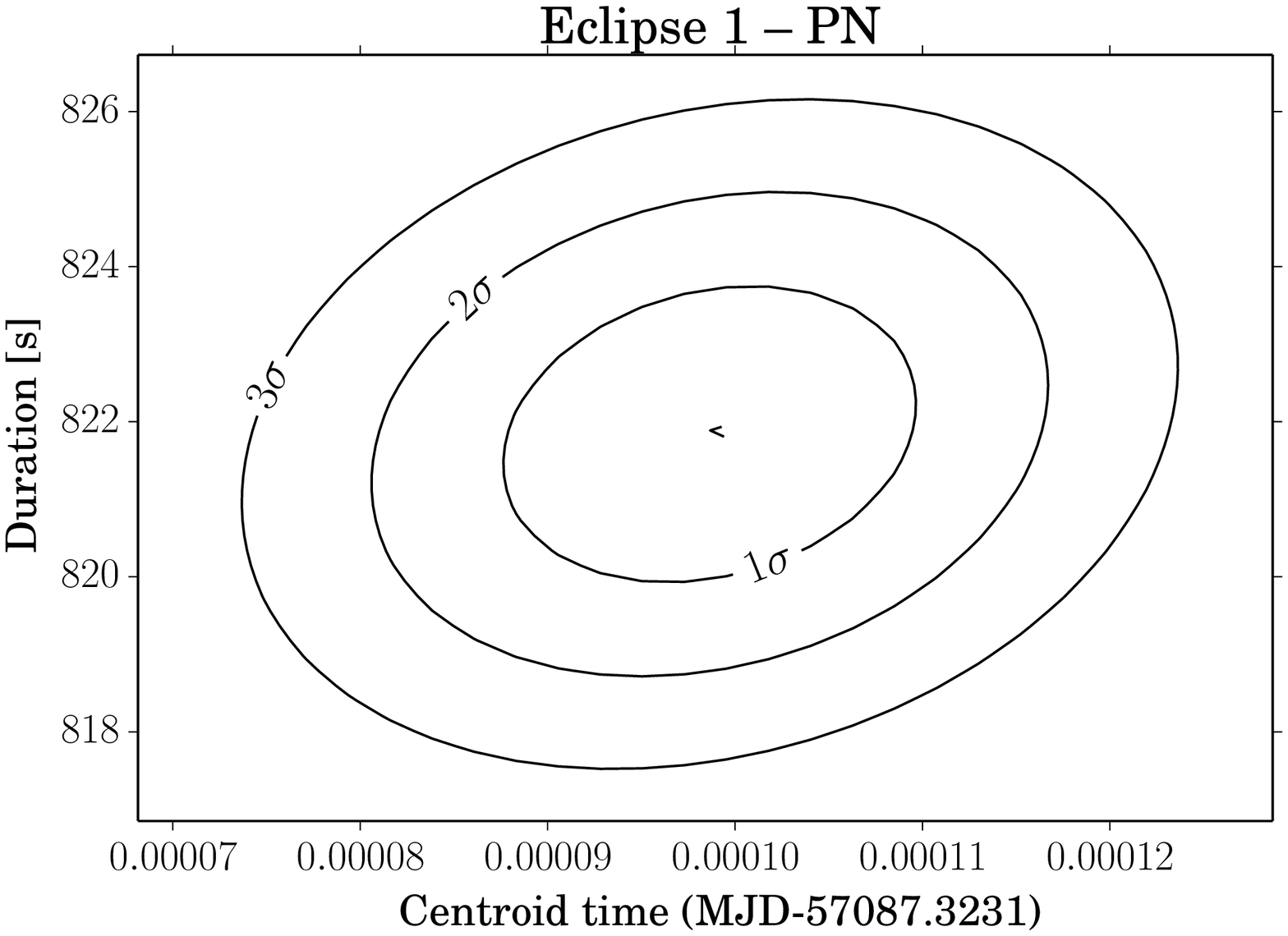}
    \includegraphics[width=9.1cm]{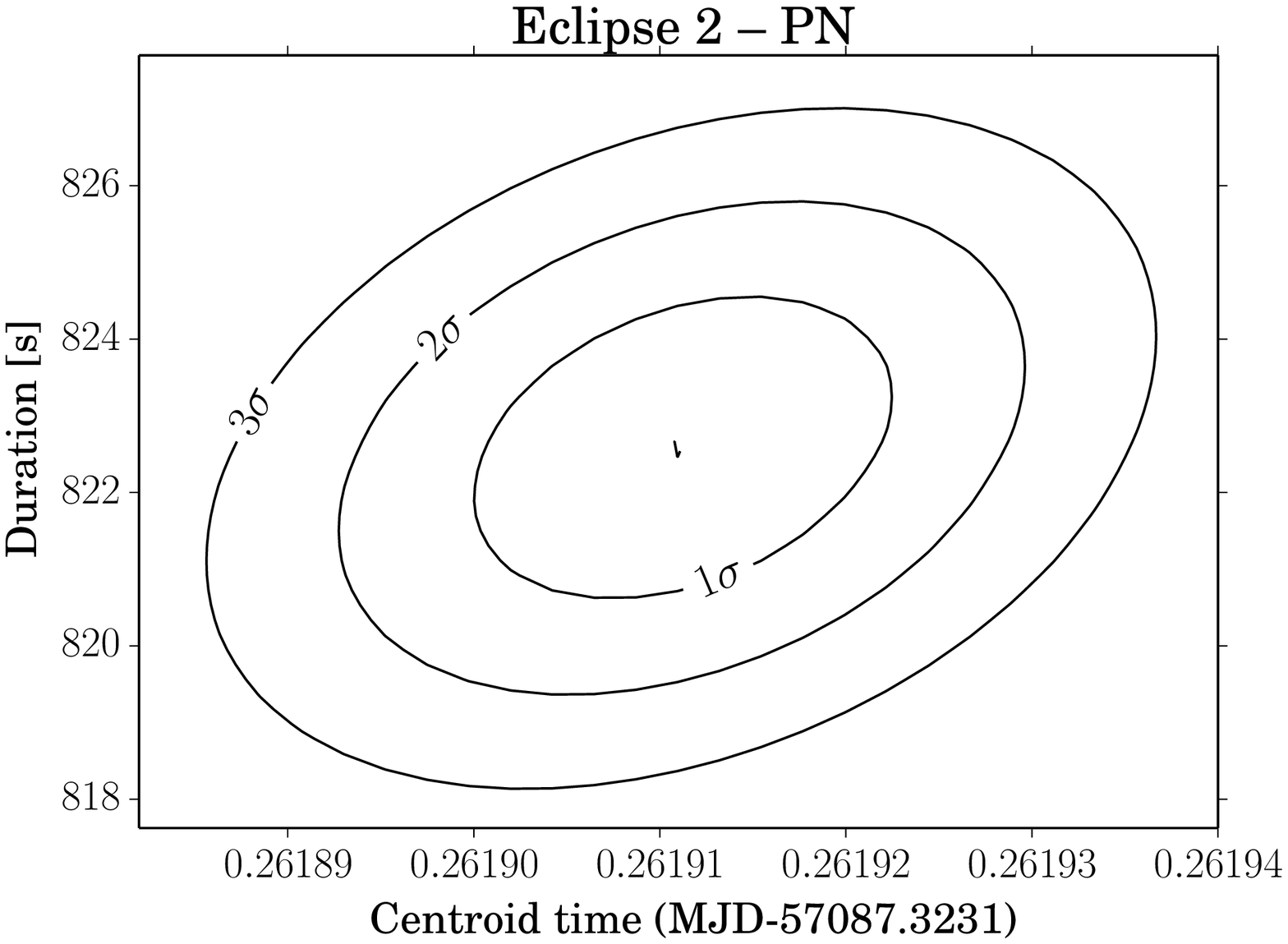}
  \caption{Contour plots of the two eclipses durations and mid-epochs obtained by fitting the 
  rectangular eclipse model to the data. The 1$\sigma$, 2$\sigma$, and 3$\sigma$ contours around 
  the best determined values are indicated.}  
  \label{fig:contours} 
\end{figure}

We used the detection of two consecutive eclipses in the 0.5-12~keV EPIC-pn lightcurve of the source 
to determine the system orbital period (we used a time bin of 20~s and verified a posteriori that a 
different choices of the time binning would not significantly affect the results). 
As the eclipses of \igr\ are virtually rectangular (see Fig.~\ref{fig:lcurve_xmm}),  
a fit to each of the two eclipses was carried out using the same technique described by \citet{bozzo07} 
in the case of the eclipsing LMXB 4U\,2129+47. 
The model we developed to fit rectangular eclipses treats the mean source count-rate outside ($F_{\rm max}$) 
and inside ($F_{\rm min}$) the eclipse as free parameters, together with the mid-eclipse epoch ($T_0$) 
and the duration of the eclipse ($D$). We performed a minimization of the 
obtained $\chi^2$ value from the fit to the data by using our dedicated IDL routine. 
The model was integrated over each time bin before computing the $\chi^2$, in order to take 
data binning into account. The routine samples the $\chi^2$ hyper-surface in a fine 
grid of values before computing the variance between the model and the data for each
set of parameters. The latter is then used to determine the local $\chi^2$ minima
in the 4D parameter space. The best fits to the two mid-eclipse epochs
were found to be $T_0(a)$=57087.32320$\pm$0.00001~MJD and 
$T_0(b)$=57087.5850110$\pm$0.00001~MJD. 
The corresponding eclipse durations were $D(a)$=821.89$^{+1.86}_{-1.95}$~s 
and $D(b)$=822.66$^{+1.89}_{-2.04}$~s. We show the relevant contour plots for the fits 
to the two eclipses in Fig.~\ref{fig:contours}. 
From these measurements, we estimated the orbital period of the system as  
$P_{\rm orb}$=22620.5$^{+2.0}_{-1.8}$~s (all uncertainties here are given at 1$\sigma$~c.l.). 
This is a slightly improved measurement compared to the value 
22623$\pm$5~s reported previously by \citet{Jaisawal15}. 

\subsection{The spectral analysis}
\label{sec:spectra}

\begin{figure}
  \includegraphics[width=5.8cm, angle=-90]{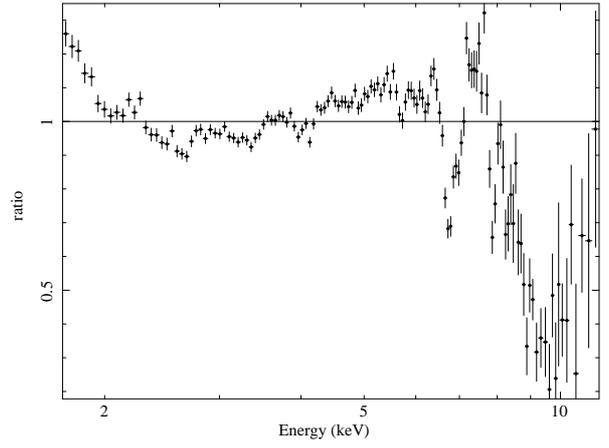}
  \caption{Ratio between the EPIC-pn data and the {\sc tbabs*diskbb} model. The plot highlights the presence of several 
  absorption features, including a broad structure around $\sim$9~keV.}  
  \label{fig:ratio} 
\end{figure}

We first extracted the source spectrum during the steady part of the \xmm\ observation, {\it i.e} filtering out the 
time intervals corresponding to the dips and the eclipses (the resulting total exposure time is 16.8~ks). 
This spectrum showed a curved shape that could be described by using a thermal component typical of an accretion disk 
({\sc diskbb} in {\sc Xspec}). We accounted for the absorption column density in the direction of the source 
by using a {\sc tbabs} component with Solar abundances. 
Although this model could describe the continuum reasonably well, the fit was far from being acceptable ($\chi^2_{\rm red}$/d.o.f.=15.94/138) 
due to the presence of several absorption features. For completeness, we mention that a fit with an absorbed power-law ({\sc 
tbabs*pow} in {\sc Xspec}) would give a significantly worse fit ($\chi^2_{\rm red}$/d.o.f.=69.75/138). 
We show in Fig.~\ref{fig:ratio} the ratio between the data and the model  
when the fit is carried out by using the {\sc tbabs*diskbb} model. 

In order to improve the fit, we first included a number of absorption Gaussian lines ({\sc gaussian} in {\sc Xspec}) to account 
for all visible features in the residuals, and then included a broad absorption feature ({\sc gabs} in {\sc Xspec}) to take into 
account the deepest structure around $\sim$9~keV (see Fig.~\ref{fig:ratio}). We fixed to zero the widths of the lines that 
could not be constrained after a preliminary fit and left this parameter free to vary in all other cases. 
This phenomenological fit was used to identify the lines that correspond to known atomic transitions and guide further analysis. 
We show all the results in Table~\ref{tab:xmmresults}, together with the identification of the lines. We marked with "?" the 
identifications of the two lines at 5.7 and 6.1~keV for which we did not find any other mention in the literature of LMXBs and that 
we thus consider as tentative associations. 

We checked for possible spectral variations by carrying out a time-resolved spectral analysis of the \xmm\ data, as well as 
an orbital-phase resolved analysis. The results of the time-resolved spectral analysis are reported 
in Table~\ref{tab:xmmresults}. 
\begin{table*}
\caption{Results of the fit to the ``steady'' emission from \igr\ recorded from \xmm\ (i.e., excluding the time intervals of the dips and 
the eclipses). The phenomenological model used here comprises a {\sc diskbb} component, plus the absorption in the direction of the source 
({\sc tbabs} in {\sc Xspec}) and all the required Gaussian absorption lines to take into account the residuals visible in Fig.~\ref{fig:ratio}. 
We also report the results of the fits carried out with the same model on the time resolved spectra extracted by dividing the steady emission 
in 5 time intervals. The latter are indicated in the first row of the table in ks from the beginning of the observation 
(see the lightcurve in Fig.~\ref{fig:lcurve_xmm}). $EW$ indicates the equivalent width of each Gaussian line. 
All uncertainties are given at 90\% c.l.}
\label{tab:xmmresults}
\scriptsize
\centering
\begin{threeparttable}
\scriptsize
\begin{tabular}{l|ccccccc}
\hline
\multicolumn{1}{c|}{\textbf{Parameter}} & \multicolumn{6}{|c}{\textbf{Value}} & \multicolumn{1}{|c}{\textbf{Identified line}}\\
\cline{2-8}
\multicolumn{1}{c|}{\textbf{(units)}} & \textbf{Steady emission} & \textbf{Dataset a} & \textbf{Dataset b} & \textbf{Dataset c} & \textbf{Dataset d} & \textbf{Dataset e} & \multicolumn{1}{|c}{\textbf{(rest energy)}}\\
\hline\\
Time interval (ks) &  & 2.05-3.75  &  8.85-12.95  & 15.45-17.05  & 28.45-34.14  & 35.15-38.65 \\
\\
$N_H$ ($10^{22}\textrm{cm}^{-2}$) & $5.6\substack{+0.3\\-0.2}$ & $5.6\pm0.2$ & $5.5\pm0.2$ & $5.1\pm0.2$ & $5.9\pm0.1$ & $5.8\substack{+0.2\\-0.1}$ \\
\\
$E_{1}$ (keV) & $2.01\substack{+0.03\\-0.05}$ & -- & -- & -- & -- & $2.04\pm0.03$ & Si XIV (2.005~keV) \\
$\sigma_{1}$ (keV) & $0.0$\tnote{$\dagger$} & -- & -- & -- & -- & $0.0$\tnote{$\dagger$} & \\
$EW_{1}$ (keV) & $-0.036\substack{+0.021\\-0.017}$ & -- & -- & -- & -- & $-0.028\pm0.010$\\
\\
$E_{2}$ (keV) & $2.52\substack{+0.03\\-0.04}$ & $2.66\substack{+0.12\\-0.13}$ & $2.52\pm0.05$ & -- & $2.56\substack{+0.06\\-0.17}$ & $2.57\substack{+0.06\\-0.07}$ & S XVI (2.62~keV) \\
$\sigma_{2}$ (keV) & $0.25\substack{+0.07\\-0.05}$ & 0.24\tnote{$\dagger$} & $0.0$\tnote{$\dagger$} & -- & $0.18\substack{+0.20\\-0.07}$ & $0.19\substack{+0.11\\-0.07}$ & \\
$EW_{2}$ (keV) & $-0.09\substack{+0.09\\-0.08}$ & $-0.09\substack{+0.04\\-0.05}$ & $-0.020\substack{+0.009\\-0.008}$ & -- & $-0.054\substack{+0.024\\-0.020}$ & $-0.075\substack{+0.027\\-0.024}$\\
\\
$E_{3}$ (keV) & $3.26\substack{+0.05\\-0.08}$ & $3.34\pm0.06$ & $3.26\pm0.05$ & -- & $3.26\pm0.09$ & $3.26\pm0.06$ & Ar XVIII (3.32~keV) \\
$\sigma_{3}$ (keV) & $0.22\substack{+0.10\\-0.09}$ & $0.0$\tnote{$\dagger$} & $0.0$\tnote{$\dagger$} & -- & $0.16\substack{+0.14\\-0.08}$ & $0.18\substack{+0.07\\-0.05}$ & \\
$EW_{3}$ (keV) & $-0.06\substack{+0.06\\-0.10}$ & $-0.015\substack{+0.010\\-0.011}$ & $-0.015\pm0.006$ & -- & $-0.026\substack{+0.022\\-0.018}$ & $-0.055\substack{+0.024\\-0.020}$ \\
\\
$E_{4}$ (keV) & $4.03\substack{+0.03\\-0.04}$ & -- & $4.08\substack{+0.06\\-0.08}$ & -- & $4.04\pm0.05$ & $4.06\substack{+0.04\\-0.05}$ & Ca XX (4.100~keV)\\
$\sigma_{4}$ (keV) & $0.15\substack{+0.05\\-0.07}$ & -- & $0.0$\tnote{$\dagger$} & -- & $0.12\substack{+0.06\\-0.07}$ & $0.0$\tnote{$\dagger$} & \\
$EW_{4}$ (keV) & $-0.040\substack{+0.032\\-0.042}$ & -- & $-0.011\pm0.007$ & -- & $-0.028\pm0.016$ & $-0.020\substack{+0.009\\-0.028}$\\
\\
$E_{5}$ (keV) & $4.8\substack{+0.1\\-0.2}$ & -- & -- & -- & -- & $4.8\substack{+0.2\\-0.1}$ & K XIX (4.797~keV) \\
$\sigma_{5}$ (keV) & $0.24\substack{+0.48\\-0.09}$ & -- & -- & -- & -- & $0.2\substack{+0.8\\-0.2}$ &  \\
$EW_{5}$ (keV) & $-0.023\substack{+0.023\\-0.051}$ & -- & -- & -- & -- & $-0.020\pm0.018$ & \\
\\
$E_{6}$ (keV) & $5.71\substack{+0.02\\-0.03}$ & -- & $5.73\pm0.04$ & $5.72\pm0.05$ & $5.74\substack{+0.03\\-0.04}$ & -- & Cr XXIII ? (5.681~keV) \\
$\sigma_{6}$ (keV) & $0.0$\tnote{$\dagger$} & -- & $0.0$\tnote{$\dagger$} & $0.0$\tnote{$\dagger$} & $0.0$\tnote{$\dagger$} & -- & \\
$EW_{6}$ (keV) & $-0.024\substack{+0.006\\-0.009}$ & -- & $-0.034\pm0.012$ & $-0.04\pm0.02$ & $-0.029\substack{+0.010\\-0.008}$ & -- & \\
\\
$E_{7}$ (keV) & $6.13\substack{+0.05\\-0.06}$ & -- & $6.10\substack{+0.02\\-0.04}$ & $6.04\substack{+0.08\\-0.11}$ & $6.18\substack{+0.09\\-0.07}$ & $6.26\substack{+0.2\\-0.5}$ & Mn XXVI ? (6.130~keV) \\
$\sigma_{7}$ (keV) & $0.16\substack{+0.11\\-0.07}$ & -- & $0.0$\tnote{$\dagger$} & $0.0$\tnote{$\dagger$} & $0.0$\tnote{$\dagger$} & $0.0$\tnote{$\dagger$} & \\
$EW_{7}$ (keV) & $-0.047\substack{+0.027\\-0.030}$ & -- & $-0.052\substack{+0.014\\-0.013}$ & $-0.030\substack{+0.018\\-0.030}$ & $-0.031\substack{+0.012\\-0.014}$ & $-0.029\pm0.015$ & \\
\\
$E_{8}$ (keV) & $6.72\pm0.01$ & $6.74\substack{+0.06\\-0.07}$ & $6.78\pm0.03$ & $6.73\pm0.04$ & $6.68\substack{+0.030\\-0.007}$ & $6.72\pm0.03$ & K$\alpha$ Fe XXV (6.70~keV)\\
$\sigma_{8}$ (keV) & $0.0$\tnote{$\dagger$} & $0.24\substack{+0.14\\-0.10}$ & $0.19\substack{+0.05\\-0.04}$ & $0.0$\tnote{$\dagger$} & $0.0$\tnote{$\dagger$} & $0.0$\tnote{$\dagger$} & \\
$EW_{8}$ (keV) & $-0.14\pm0.03$ & $-0.19\pm0.06$ & $-0.24\pm0.04$ & $-0.10\substack{+0.02\\-0.07}$ & $-0.12\pm0.02$ & $-0.11\pm0.02$\\
\\
$E_{9}$ (keV) & $7.00\pm0.03$ & -- & -- & -- & $6.94\pm0.03$ & $6.99\substack{+0.03\\-0.06}$ & K$\alpha$ Fe XXVI (6.97~keV) \\
$\sigma_{9}$ (keV) & $0.0$\tnote{$\dagger$} & -- & -- & -- & $0.0$\tnote{$\dagger$} & $0.0$\tnote{$\dagger$} & \\
$EW_{9}$ (keV) & $-0.08\pm0.02$ & -- & -- & -- & $-0.11\pm0.02$ & $-0.068\substack{+0.028\\-0.015}$\\
\\
$E_{10}$ (keV) & $7.86\substack{+0.02\\-0.01}$ & $7.87\substack{+0.04\\-0.06}$ & $7.87\substack{+0.05\\-0.04}$ & $7.89\pm0.06$ & $7.87\pm0.07$ & -- & K$\beta$ Fe XXV (7.88~keV)\\
$\sigma_{10}$ (keV) & $0.0$\tnote{$\dagger$} & $0.0$\tnote{$\dagger$} & $0.0$\tnote{$\dagger$} & $0.0$\tnote{$\dagger$} & $0.0$\tnote{$\dagger$} & -- & \\
$EW_{10}$ (keV) & $-0.092\substack{+0.019\\-0.026}$ & $-0.13\substack{+0.06\\-0.05}$ & $-0.11\pm0.04$ & $-0.11\substack{+0.04\\-0.08}$ & $-0.068\pm0.033$ & -- \\
\\
$E_{\textrm{gabs}}$ (keV) & $9.5\substack{+0.4\\-0.2}$ & $9.2\substack{+1.8\\-0.2}$ & $9.6\substack{+0.9\\-0.3}$ & $9.7$\tnote{$\dagger$} & $9.3\substack{+0.2\\-0.1}$ & $9.7$\tnote{$\dagger$}\\
$\sigma_{\textrm{gabs}}$ (keV) & $0.8\substack{+0.3\\-0.2}$ & $0.6\substack{+0.7\\-0.2}$ & $0.8\substack{+0.5\\-0.2}$ & $2.1\substack{+0.6\\-0.4}$ & $0.7\pm0.2$ & $1.0\pm0.1$\\
$S_{\textrm{gabs}}$ & $3.0\substack{+1.2\\-0.9}$ & $2.2\substack{+9.6\\-0.6}$ & $2.9\substack{+0.9\\-0.6}$ & $6.3\substack{+2.7\\-1.6}$ & $2.4\substack{+0.7\\-0.5}$ & $3.5\pm0.8$\\
\\
$kT_{\rm diskbb}$ (keV) & $1.27\substack{+0.05\\-0.13}$ & $1.32\substack{+0.03\\-0.04}$ & $1.38\pm0.04$ & $1.7\substack{+0.9\\-0.3}$ & $1.27\pm0.03$ & $1.24\pm0.03$\\
$N_{\rm diskbb}$ & $9.9\substack{+40.5\\-2.0}$ & $7.8\pm1.5$ & $6.4\substack{+1.1\\-1.0}$ & $3.0\substack{+1.4\\-1.6}$ & $9.9\substack{+1.6\\-1.1}$ & $11.0\substack{+2.0\\-0.8}$\\
\\
$_{F_{0.5--10\textrm{keV}}}$ ($10^{-10}\frac{\textrm{erg}}{\textrm{cm}^2\textrm{s}}$) & $1.563\substack{+0.003\\-0.203}$ & $1.55\substack{+0.02\\-0.04}$ & $1.6\substack{+0.1\\-0.5}$ & $1.6\substack{+0.1\\-0.9}$ & $1.55\substack{+0.1\\-0.4}$ & $1.55\substack{+0.04\\-0.01}$\\
\\
$\chi^2/\nu$ (d.o.f) & $1.41$ ($101$) & $1.09$ ($100$) & $1.42$ ($93$) & $1.08$ ($98$) & $1.07$ ($101$) & $1.14$ ($98$)\\
\\
Exposure time (s) & 16780 & 1675 &  3654 & 1580 & 5418 & 3456 \\
\hline
\end{tabular}
\begin{tablenotes}\footnotesize
 \item[$\dagger$] parameter frozen in the fit. 
\end{tablenotes}
\end{threeparttable}
\end{table*}
\begin{table*}[tt!]
\centering
\caption{Same as Table~\ref{tab:xmmresults} but for the orbital phase resolved spectral analysis. We repeated here the first column in order to 
facilitate the comparison between the results of the averaged steady emission and those obtained from the phase resolved spectral analysis. 
We excluded from the orbital phase resolved spectral analysis the phases corresponding to the eclipses and the dips (see text for details).}
\tiny
\label{tab:xmmresults2}
\begin{threeparttable}
\scriptsize
\begin{tabular}{l|cccc}
\hline
\multicolumn{1}{c|}{\textbf{Parameter}} & \multicolumn{4}{|c}{\textbf{Value}}\\
\cline{2-5}
\multicolumn{1}{c|}{\textbf{(units)}} & \textbf{Steady emission} & \textbf{Phase $0.0-0.2$} & \textbf{Phase $0.2-0.4$} & \textbf{Phase $0.4-0.6$}\\
\hline\\
$N_H$ ($10^{22}\textrm{cm}^{-2}$) & $5.6\substack{+0.3\\-0.2}$ & $5.6\pm0.1$ & $5.5\pm0.2$ & $5.5\substack{+0.2\\-0.3}$\\
\\
$E_{1}$ (keV) & $2.01\substack{+0.03\\-0.05}$ & $2.07\substack{+0.03\\-0.06}$ & $2.09\substack{+0.04\\-0.03}$ & $2.00\substack{+0.03\\-0.02}$\\
$\sigma_{1}$ (keV) & $0.0$\tnote{$\dagger$} & $0.0$\tnote{$\dagger$} & $0.0$\tnote{$\dagger$} & $0.0$\tnote{$\dagger$}\\
$EW_{1}$ (keV) & $-0.036\substack{+0.021\\-0.017}$ & $-0.014\pm0.08$ & $-0.016\substack{+0.011\\-0.007}$ & $-0.026\pm0.008$\\
\\
$E_{2}$ (keV) & $2.52\substack{+0.03\\-0.04}$ & $2.52\pm0.06$ & $2.54\substack{+0.03\\-0.02}$ & $2.4\substack{+0.2\\-0.3}$\\
$\sigma_{2}$ (keV) & $0.25\substack{+0.07\\-0.05}$ & $0.18\substack{+0.06\\-0.04}$ & $0.12\substack{+0.05\\-0.04}$ & $0.5\substack{+0.2\\-0.1}$\\
$EW_{2}$ (keV) & $-0.09\substack{+0.09\\-0.08}$ & $-0.048\pm0.015$ & $-0.047\pm0.013$ & $-0.17\substack{+0.11\\-0.06}$\\
\\
$E_{3}$ (keV) & $3.26\substack{+0.05\\-0.04}$ & -- & $3.21\pm0.06$ & $3.37\substack{+0.04\\-0.06}$\\
$\sigma_{3}$ (keV) & $0.22\substack{+0.10\\-0.09}$ & -- & $0.21\substack{+0.14\\-0.09}$ & $0.0$\tnote{$\dagger$}\\
$EW_{3}$ (keV) & $-0.06\substack{+0.06\\-0.10}$ & -- & $-0.050\substack{+0.020\\-0.019}$ & $-0.011\pm0.06$\\
\\
$E_{4}$ (keV) & $4.03\substack{+0.03\\-0.04}$ & $4.06\substack{+0.07\\-0.10}$ & $4.08\substack{+0.04\\-0.07}$ & $4.04\substack{+0.03\\-0.06}$\\
$\sigma_{4}$ (keV) & $0.15\substack{+0.05\\-0.07}$ & $0.0$\tnote{$\dagger$} & $0.0$\tnote{$\dagger$} & $0.0$\tnote{$\dagger$}\\
$EW_{4}$ (keV) & $-0.040\substack{+0.032\\-0.042}$ & $-0.012\pm0.05$ & $-0.014\pm0.07$ & $-0.017\pm0.007$\\
\\
$E_{5}$ (keV) & $4.8\substack{+0.1\\-0.2}$ & -- & -- & --\\
$\sigma_{5}$ (keV) & $0.24\substack{+0.48\\-0.09}$ & -- & -- & --\\
$EW_{5}$ (keV) & $-0.023\substack{+0.023\\-0.051}$ & -- & -- & --\\
\\
$E_{6}$ (keV) & $5.71\substack{+0.02\\-0.03}$ & $5.67\pm0.04$ & $5.78\substack{+0.03\\-0.05}$ & $5.92\pm0.06$\\
$\sigma_{6}$ (keV) & $0.0$\tnote{$\dagger$} & $0.0$\tnote{$\dagger$} & $0.0$\tnote{$\dagger$} & $0.0$\tnote{$\dagger$}\\
$EW_{6}$ (keV) & $-0.024\substack{+0.006\\-0.009}$ & $-0.021\pm0.08$ & $-0.029\substack{+0.010\\-0.009}$ & $-0.022\pm0.010$\\
\\
$E_{7}$ (keV) & $6.13\substack{+0.05\\-0.06}$ & $6.20\pm0.06$ & $6.18\pm0.04$ & --\\
$\sigma_{7}$ (keV) & $0.16\substack{+0.11\\-0.07}$ & $0.0$\tnote{$\dagger$} &  $0.0$\tnote{$\dagger$} & --\\
$EW_{7}$ (keV) & $-0.047\substack{+0.027\\-0.030}$ & $-0.022\pm0.012$ & $-0.032\substack{+0.011\\-0.012}$ & --\\
\\
$E_{8}$ (keV) & $6.72\pm0.01$ & $6.68\pm0.02$ & $6.71\pm0.02$ & $6.73\substack{+0.02\\-0.03}$\\
$\sigma_{8}$ (keV) & $0.0$\tnote{$\dagger$} & $0.0$\tnote{$\dagger$} & $0.0$\tnote{$\dagger$} & $0.0$\tnote{$\dagger$}\\
$EW_{8}$ (keV) & $-0.14\pm0.03$ & $-0.12\pm0.02$ & $-0.11\substack{+0.01\\-0.02}$ & $-0.11\pm0.02$\\
\\
$E_{9}$ (keV) & $7.00\pm0.03$ & $6.96\pm0.03$ & $6.99\substack{+0.06\\-0.04}$ & $6.99\substack{+0.06\\-0.04}$\\
$\sigma_{9}$ (keV) & $0.0$\tnote{$\dagger$} & $0.0$\tnote{$\dagger$} & $0.0$\tnote{$\dagger$} & $0.0$\tnote{$\dagger$}\\
$EW_{9}$ (keV) & $-0.08\pm0.02$ & $-0.11\pm0.02$ & $-0.11\pm0.02$ & $-0.064\pm0.022$\\
\\
$E_{10}$ (keV) & $7.86\substack{+0.02\\-0.01}$ & $7.86\pm0.05$ & $7.89\pm0.04$ & $7.85\substack{+0.04\\-0.03}$\\
$\sigma_{10}$ (keV) & $0.0$\tnote{$\dagger$} & $0.0$\tnote{$\dagger$} & $0.0$\tnote{$\dagger$} & $0.0$\tnote{$\dagger$}\\
$EW_{10}$ (keV) & $-0.092\substack{+0.019\\-0.026}$ & $-0.075\substack{+0.029\\-0.024}$ & $-0.10\pm0.03$ & $-0.11\pm0.03$\\
\\
$E_{\textrm{gabs}}$ (keV) & $9.5\pm0.2$ & $9.4\substack{+0.5\\-0.2}$ & $9.6\substack{+0.9\\-0.4}$ & $9.6\substack{+0.9\\-0.4}$\\
$\sigma_{\textrm{gabs}}$ (keV) & $0.8\substack{+0.3\\-0.2}$ & $0.9\pm0.2$ & $0.7\substack{+0.5\\-0.2}$ & $0.7\substack{+0.5\\-0.2}$\\
$S_{\textrm{gabs}}$ & $3.0\substack{+1.2\\-0.9}$ & $3.1\substack{+1.5\\-0.7}$ & $3.3\substack{+12.8\\-1.4}$ & $3.3\substack{+12.3\\-1.4}$\\
\\
$kT_{\rm diskbb} (keV)$ & $1.25\substack{+0.05\\-0.13}$ & $1.32\pm0.02$ & $1.26\substack{+0.03\\-0.02}$ & $1.26\pm0.03$\\
$N_{\rm diskbb}$ & $9.9\substack{+40.5\\-2.0}$ & $7.3\pm0.6$ & $10.2\substack{+1.2\\-1.0}$ & $10.2\substack{+1.2\\-1.0}$\\
\\
$_{F_{0.5--10\textrm{keV}}}$ ($10^{-10}\frac{\textrm{erg}}{\textrm{cm}^2\textrm{s}}$) & $1.563\substack{+0.002\\-0.203}$ & $1.43\pm0.01$ & $1.55\substack{+0.03\\-0.04}$ & $1.55\pm0.03$\\
$\chi^2/\nu$ (d.o.f) & $1.41$ ($101$) & $1.27$ ($104$) & $1.34$ ($103$) & $1.34$ ($103$)\\
\\
Exposure time (s) &  16780 & 6564 & 6478 & 4943 \\\hline

\end{tabular}
\begin{tablenotes}\footnotesize
 \item[$\dagger$] parameter frozen in the fit. 
\end{tablenotes}
\end{threeparttable}
\end{table*}
The results obtained from the orbital phase resolved spectral analysis are reported in Table~\ref{tab:xmmresults2}. 
In both cases we could not find evidence of significant spectral variability. 
We also noticed that, owing to the reduced statistics of these spectra, obtaining meaningful fits in the 6-8~keV range was 
particularly challenging due to the blending of several lines.  

\begin{figure}
    \includegraphics[width=5.9 cm,angle=-90]{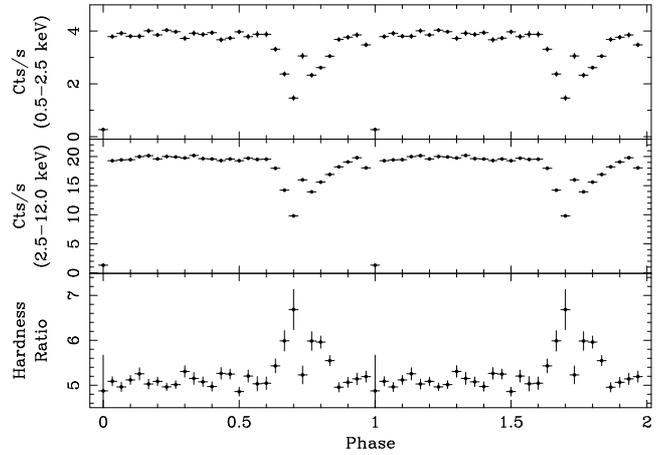}
  \caption{EPIC-pn lightcurve of \igr\ folded on the best orbital period determined in this work, 
  $P_{\rm orb}$=22620.5$^{+2.0}_{-1.8}$~s (see Sect.~\ref{sec:eclipses}). 
  The upper (middle) panel shows the folded source lightcurve in the soft (hard) X-ray energy band, while the bottom panel shows the 
  HR (i.e. the ratio of the source count-rate in the hard, 2.5–12~keV, and soft, 0.5–2.5~keV, energy bands).}   
  \label{fig:xmmorbital}  
\end{figure}

The spectral analysis of the X-ray dips was carried out separately, as the hardness ratio (HR) calculated from the soft (0.5-2.5~keV) 
and hard (2.5-12.0~keV) X-ray lightcurves of the source as a function of the orbital phase provided clear evidence for spectral variability 
during these events (see Fig.~\ref{fig:xmmorbital}). 
We extracted orbital phase resolved spectra during phases corresponding to the HR variations: 0.61-0.65, 0.65-0.68, 0.68-0.75, 0.75-0.82, 
and 0.82-0.85. In all cases a reasonable good fit ($\chi_{\rm red}^2$=0.9-1.2) could 
be obtained with an absorbed {\sc diskbb} model. Owing to the poor statistics of these spectra 
(integration times of few ks) we could not measure significant changes in the temperature and radius of the {\sc diskbb} component, 
neither in the absorption column density. Furthermore, several absorption lines appeared as blended and thus we were unable to 
constrain within a reasonable accuracy their parameters in the fits. In order to clarify the origin of the HR increase close to the dips 
(as visible in Fig.~\ref{fig:xmmorbital}), we extracted a single spectrum by combining orbital phases between 0.65 and 0.82. 
This spectrum revealed a modest (but significant) increase in the column density of both the neutral and the additional ionized absorber. 
This is discussed more in Sect.~\ref{sec:spectraph}. 

We also extracted the source spectrum during the two eclipses. Since in this case the source flux was significantly lower, we also made 
use of the spectra extracted from the two MOS cameras and fit them together with the EPIC-pn (the integration time was too short 
to obtain meaningful spectra from the two RGSs and thus we did not include these data in the combined fit). The spectrum could be 
equally well fit with an absorbed {\sc diskbb} ($N_{\rm H}$=(5.3$\pm$0.6)$\times$10$^{22}$~cm$^{-2}$, 
$kT$=1.17$\pm$0.09, $N_{\rm diskbb}$=0.8$\pm$0.4, 
$\chi_{\rm red}^2$/d.o.f.=0.92/98) or a power-law model ($N_{\rm H}$=(8.8$\pm$0.9)$\times$10${22}$~cm$^{-2}$, 
$\Gamma$=3.9$\pm$0.3, $\chi_{\rm red}^2$/d.o.f.=0.98/98). The measured 0.5--10~keV 
X-ray flux was 9.5$\times$10$^{-12}$~erg~cm$^{-2}$~s$^{-1}$. We introduced in the fit inter-calibration constants 
between the three instruments and found them to be compatible with unity (to within the uncertainties).

\subsubsection{A physical spectral model}
\label{sec:spectraph}

\begin{figure}
    \includegraphics[width=6.1  cm,angle=-90]{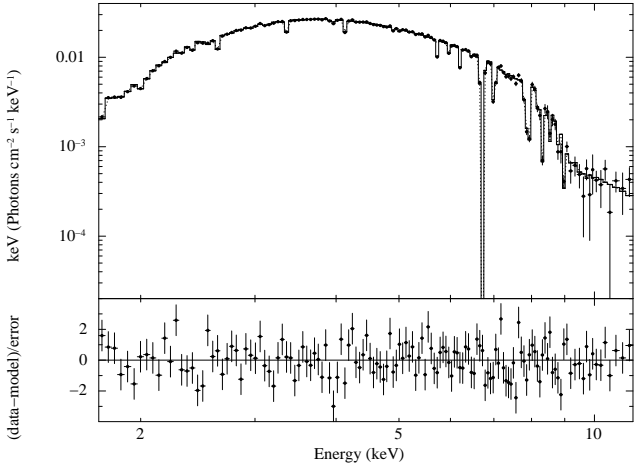}
  \caption{Best fit to the average EPIC-pn spectrum of the steady emission from \igr\ (excluding the time intervals 
  corresponding to the eclipses and the dips) with the physical model described in Sect.~\ref{sec:spectraph}. 
  We show here the unfolded spectrum. 
  The residuals from the fit are shown in the bottom panel.}   
  \label{fig:bestfit}  
\end{figure}

The spectral properties of the source analyzed in Sect.~\ref{sec:spectra}, together with the characteristics of the lightcurve  
discussed in Sect.~\ref{sec:eclipses}, are strongly reminiscent of high inclination LMXBs with absorbing/outflowing material 
located above the accretion disk (see Sect.~\ref{sec:discussion}). In order to provide a more physical description of the spectral energy 
distribution of \igr,\ we made use of a model comprising a {\sc diskbb} component which is affected by both the Galactic absorption 
and a local ``warm'' absorber ({\sc warmabs}\footnote{http://heasarc.gsfc.nasa.gov/xstar/docs/html/node102.html} 
in {\sc Xspec}). This model takes into account the ionization status of the gas, 
its turbulent broadening velocity, its bulk velocity (through the measurement of redshifted or blueshifted lines), and the abundance 
of the different materials with respect to Solar values. 
Following the identification of several absorption lines 
in Sect.~\ref{sec:spectra}, we left free to vary in the fit the abundances corresponding to Si, S, Ar, Ca, Mn, K, and Fe. 
A {\sc cabs}  component, accounting for Compton scattering (not included in {\sc warmabs}), was also added to the fit. We fixed the 
column density of this component to be the same of the {\sc warbmabs} multiplied by a factor of 1.21 to take roughly into account 
the number of electrons per hydrogen atom \citep[in case of Solar abundances; see, e.g.][and references therein]{trigo14}. 
We changed the default ionizing continuum in {\sc warmabs}, that is a power-law of photon index $\Gamma$=2, with a blackbody 
ionizing continuum with kT$\sim$1.2~keV. Even though the source might undergo some spectral state 
transitions characterized by a non thermal power-law like component, this does not seem the case 
for most of the outburst (see Sect.~\ref{sec:integral}). As it is not possible to select a {\sc diskbb} component as a ionizing continuum 
in the {\sc warmabs} model, we considered that the available blackbody continuum with a similar temperature was 
a reasonably good approximation in the case of \igr\ \citep[as also done in other sources, see e.g.][]{trigo13}. 
We also assumed 10$^{12}$~cm$^{-3}$ for the ionized medium density, a value typically used in LMXBs \citep[see, e.g.,][]{trigo14}.  
The drop in the source X-ray emission at energies $\gtrsim$9~keV 
that required the addition of a {\sc gabs} component in the phenomenological model was completely 
accounted for by the addition of the {\sc warmabs} model. We included in the fit an additional edge at $\sim$8.2~keV to minimize 
the residuals still visible around this energy and two absorption Gaussian lines with centroid 
energies of $\sim$5.7~keV and $\sim$6.1~keV. The latter improved the fit to the Cr and Mn lines, which were not fully taken into account 
by the {\sc warmabs} component. This model provided an acceptable result (see Table~\ref{tab:bestfit}). 
The residuals left from this fit did not suggest the presence of additional spectral components, even though some structures remained 
visible close to the iron lines at 6.7-7.0~keV (see Fig.~\ref{fig:bestfit}). 
Such structures might result from a non-optimal description of the complex absorbing medium in \igr\ through the 
{\sc warmabs} model, e.g. due to the presence of different broadening or outflowing velocities. 

We used the best fit model also to characterize the spectrum extracted during the orbital phases corresponding to the dips and the most 
evident HR variations (0.65-0.72 and 0.72-0.82, see Sect.~\ref{sec:spectra}). This allowed a more physical comparison between the average 
emission of the source and the emission during the dips. Owing to the reduced statistics of the dip spectrum, we fixed in the fit 
those parameters that could not be constrained to the values determined from the average spectrum. The results of this fit, summarized 
in Table~\ref{tab:bestfit}, suggest that the dips are likely caused by a moderate (but significant) increase in the 
column density of the neutral and ionized absorber close to the source. 
\begin{table}
\caption{Spectral parameters obtained from the fit of the physical model in Sect.~\ref{sec:spectraph} to the averaged and dip 
EPIC-pn spectra of \igr\ (see text for details). The model in {\sc Xspec} is 
{\sc tbabs*cabs*edge*warmabs}{\sc *(gaussian+gaussian+diskbb)}. Here,   
$N_H^{\rm warm}$ is the column density of the warm absorbing material, $\log(\xi)$ its ionization status, 
$\sigma_{\rm v}$ its turbulent velocity broadening, $v$ the bulk velocity shift, and $EW$ the equivalent width of the 
Gaussian lines. The abundance of different 
elements is indicated as a fraction of the corresponding Solar value.}
\label{tab:bestfit}
\centering
\begin{threeparttable}
\scriptsize
\centering
\begin{tabular}{l|ll}
\hline
\multicolumn{1}{c|}{\textbf{Parameter}} & \multicolumn{2}{|c}{\textbf{Value}}\\
\cline{2-3}
\multicolumn{1}{c|}{\textbf{(units)}} & \multicolumn{1}{c}{\textbf{Steady emission}} & \textbf{Dips}\\
\hline\\
{\sc tbabs}\\
$N_H$ ($10^{22}\textrm{cm}^{-2}$) & $5.42\substack{+0.09\\-0.09}$ & $5.86\substack{+0.08\\-0.29}$ \\
\hline\\
{\sc edge}\\
$E_{edge}$ (keV) & $8.14\substack{+0.06\\-0.03}$ & 8.14$\tnote{$\dagger$}$\\
$\tau$  & $0.45\substack{+0.07\\-0.09}$ & 0.44$\tnote{$\dagger$}$\\
\hline\\
{\sc warmabs}\\
$N_H^{\rm warm}$ ($10^{22}\textrm{cm}^{-2}$) & $113\substack{+30\\-19}$  & $160\substack{+9\\-10}$\\
$\log(\xi)$ (erg~cm~s$^{-1}$) & $4.02\substack{+0.03\\-0.03}$  & $3.87\substack{+0.08\\-0.08}$\\
Si & $1.96\substack{+0.23\\-0.37}$  & $3.2\substack{+0.9\\-0.5}$\\
S  & $0.58\substack{+0.19\\-0.11}$ & $4.0\substack{+1.0\\-1.0}$\\
Ar & $1.14\substack{+0.37\\-0.25}$ & $0.1\substack{+0.4\\-0.1}$\\
Ca & $1.17\substack{+0.37\\-0.28}$  & $0.4\substack{+0.6\\-0.3}$\\
K & 0$\tnote{$\dagger\dagger$}$ & 0$\tnote{$\dagger$}$\\
Cr & $0.77\substack{+0.17\\-1.00}$ & $0.3\substack{+1.0\\-0.3}$\\
Mn & $0.57\substack{+0.56\\-0.30}$ & $1.0\substack{+2.8\\-1.0}$\\
Fe & $1.74\substack{+0.14\\-0.11}$ & $1.4\substack{+0.2\\-0.2}$\\
$\sigma_{\rm v}$ (km~s$^{-1}$) &  $290\substack{+39\\-42}$ &  $295\substack{+42\\-10}$\\
$v$ (km~s$^{-1}$) &  $2231\substack{+60\\-162}$ &  $1615\substack{+92\\-34}$\\
\hline\\
{\sc Gaussian}\\
$E_{6}$ (keV) & $6.20\substack{+0.04\\-0.03}$ & 6.20$\tnote{$\dagger$}$ \\
$\sigma_{6}$ (keV) & 0$\tnote{$\dagger$}$ & 0$\tnote{$\dagger$}$\\ 
$EW_{6}$ (keV) & $-0.020\substack{+0.009\\-0.008}$ & $-0.015\substack{+0.009\\-0.007}$\\
\hline\\
{\sc Gaussian}\\
$E_{7}$ (keV) & $5.71\substack{+0.02\\-0.03}$ & 5.71$\tnote{$\dagger$}$ \\
$\sigma_{7}$ (keV) & 0$\tnote{$\dagger$}$ & 0$\tnote{$\dagger$}$ \\
$EW_{7}$ (keV) & -0.017$\pm$0.007 & $-0.017\substack{+0.003\\-0.005}$\\ 
\hline\\
{\sc diskBB}\\
$kT_{\rm diskbb} (keV)$ & $1.252\substack{+0.004\\-0.004}$ & 1.21$\pm$0.01\\
$N_{\rm diskbb}$ & $11.49\substack{+0.04\\-0.16}$ & $12.6\substack{+1.4\\-1.5}$\\
\hline\\
$_{F_{0.5--10\textrm{keV}}}$ & $1.564\substack{+0.033\\-0.011}$ & $1.17\substack{+0.02\\-0.02}$\\
(Unabsorbed flux$\tnote{$\dagger\dagger\dagger$}$) & (5.6) &  (5.3) \\
$\chi^2/\nu$ (d.o.f) & $1.32$ ($121$) & $0.99$ ($107$) \\
\hline 
\end{tabular}
\begin{tablenotes}\footnotesize
\item[$\dagger$] parameter frozen in the fit.
\item[$\dagger\dagger$] the fit turned out to be insensitive to sufficiently small values of this parameter. Therefore, we fixed it to zero. 
\item[$\dagger\dagger\dagger$] Absorbed and unabsorbed fluxes are given in units of $10^{-10}$~erg~cm$^{-2}$~s$^{-1}$ in the 0.5--10~keV energy range 
estimated from {\sc Xspec} by setting the {\sc tbabs} and {\sc warmabs} absorption column densities to zero. 
\end{tablenotes}
\end{threeparttable}
\end{table}

\subsection{Timing analysis}
\label{sec:timing}

We also performed a timing analysis of the EPIC-pn data to search for coherent signals. 
We first reported all photons arrival times to the Solar system barycentre using the Chandra 
position of the source (see Sect.~\ref{sec:intro}). The orbital motion of the compact object limits the 
length of the time series over which a search for periodicities can be performed without taking 
into account the Doppler shift induced by the motion of the source. To evaluate such a time interval 
we calculated Eq.~21 of \citet{johnston11} for an orbital period of $P_{\rm orb}=6.28$~hr and an 
inclination $i\sim80^{\circ}$, obtaining:
\begin{equation}
\Delta t^*_{best}=448.6 \left(\frac{\nu}{300\,\mbox{Hz}} \right)^{-1/2} \left(\frac{4}{m} 
\right)^{1/2} \left(\frac{M_2}{0.3 M_{\odot}} \right)^{-2/5} s.
\end{equation}
Assuming the case of a signal with a  putative frequency of $\nu = 300$ Hz, described by $m = 1$ 
harmonic and emitted by an object orbiting around a $M_2 \sim 0.3 M_{\odot}$ companion star 
(Zdziarski et al. 2016, in preparation), we 
then calculated a Fourier power spectrum in intervals of length $\Delta t_{\rm best}=448.6$~s. The 
time series was sampled at eight times the minimum time resolution of the timing mode of the EPIC-pn 
($t_{\rm res}=2.952\times10^{-5}$ s), obtaining a value of the Nyquist frequency, $\nu_{\rm Ny}=(8t_{\rm res})^{-1}/2=2117$~Hz. 
We then inspected frequencies up to $\nu_{\rm Ny}$ (with a frequency resolution 
of $\delta \nu=1/\Delta t^*_{\rm best}=2.2\times10^{-3}$ Hz) to search for a coherent periodic signal, 
but found none significant with an upper limit on the pulse amplitude of $A<12.5\%$ (3$\sigma$ c.l.). 

In order to improve the sensitivity to pulsed signals emitted by a source in a binary system, 
we used a quadratic coherence recovery technique (QCRT), following the guidelines given by 
\citet{wood91} and \citet{vaughan94}. Such a technique is based on the correction of 
the photon arrival times with a quadratic transformation, under the assumption that the 
sinusoidal Doppler modulation introduced by the orbital motion can be approximated by a 
parabola over a time interval short enough. To estimate such a length we calculated Eq.~22 of 
\citet{johnston11} for the same parameters mentioned above:
\begin{equation}
\Delta t_{best}= 1636 \left(\frac{\nu}{300\,\mbox{Hz}} \right)^{-1/3} \left(\frac{4}{m} 
\right)^{1/3} \left(\frac{M_2}{0.3 M_{\odot}} \right)^{-4/15} s.
\end{equation}
In order to increase the speed of the power spectrum calculation, we considered a value of the interval 
length equal to a power of two of the time resolution of the time series, $\Delta t_{\rm best}=2^26 t_{\rm res}=1981$, 
and divided the time series  into $N_{\rm intv} = 20$ intervals. Before performing a signal 
search, the arrival times of the X-ray photons in each of these intervals were corrected 
using a quadratic relation  $t' = \alpha t^2$. The parameter $\alpha$ was varied in steps 
equal to $\delta \alpha = 1/(2\nu_{\rm Ny} \Delta t^2_{\rm best}) = 6\times10^{-11} s^{-1}$, into a 
range determined by consideration on the possible size of the two components. Calculating 
the expression given by \citet{wood91} for $P_{\rm orb} = 6.28$~hr, we considered 
$\alpha_{\rm max} = −\alpha_{\rm min} = 1.54 \times 10^{-7} [M_2/(M_1 + M_2)] [(M_1 + M_2)/M_{\odot}]^{1/3} s^{-1}$. 
Assuming as example values $M_1 = 1.7 M_{\odot}$, and $M_2 = 0.3 M_{\odot}$, we obtained 
$\alpha_{\rm max} = −\alpha_{\rm min} = 2.9 \times 10^{-8}$~s. The number of preliminary arrival times correction 
made on each of the $N_{\rm intv} = 20$ time intervals, is then $N_\alpha = 2\alpha_{\rm max}/\delta\alpha = 967$. 
A power spectrum was produced for each of these trial corrections, inspecting frequencies up 
to $\nu_{\rm Ny}=(8t_{\rm res})^{-1}/2=2117$~Hz with a resolution of 
$\delta \nu=1/\Delta t_{\rm best}=1/(2^26 t_{\rm res})=5\times10^{-4}$~Hz, giving a total number of frequencies 
trials equal to $N_{\nu}=\nu_{\rm Ny}/\delta{\nu}=2^{22}$. The total number of trials is then 
$N_{\rm intv}N_{\alpha}N_{\nu}=8.1\times10^{10}$, and assuming that in the absence of any signal 
the noise in the power spectrum is distributed as a $\chi^2$ with two degrees of freedom, 
we then set a post-trial detection level at 3$\sigma$ confidence level of $P_{\rm det} = 62.1$. 
No signal was significantly detected and upper limits of $\simeq5\%$ (3$\sigma$ c.l.) 
were typically found in the different time intervals.

\section{Discussion}
\label{sec:discussion}

We reported on the analysis of the X-ray data collected during the first 
outburst of the \inte\ transient \igr.\ The source displayed a relatively soft X-ray emission during 
most of the event (lasting about 9 months) and thus it was not significantly detected by the hard X-ray imager 
IBIS/ISGRI on-board \inte.\ The only exception is the time interval corresponding to the satellite 
revolution 1458 where the source underwent a possible spectral state transition and its X-ray emission 
was recorded up to $\sim$100~keV. During the \inte\ revolution 1458, the spectrum of the source could be well 
described by using a relatively hard power-law, with a photon index of $\Gamma\sim2$. During all the remaining outburst 
phases covered by \inte\ the source was only significantly detected in by JEM-X at energies $\lesssim$10~keV. 
The average JEM-X spectrum, however, was characterized by a low statistics and did not permit a clear distinction between different 
possible spectral models. The JEM-X data showed, nevertheless, that the X-ray flux of the source remained relatively stable 
in the 3-10~keV energy band and compatible with that recorded by \swift\,/XRT before the beginning of the outburst decay.   

The \swift\,/XRT observations covered a large part of the source outburst, excluding the time period 2014 November to 2015 
February when the visibility of the Galactic center was limited by Sun constraints. XRT data did not reveal any dramatic spectral 
change in the X-ray emission from the source during the outburst. Its spectrum was dominated by a thermal component, interpreted 
as being due to the presence of an accretion disk. The hard X-ray emission recorded during the \inte\ 
revolution 1458 could thus have been due to a temporary spectral state transition of the source, similarly to what is often 
observed during the outbursts of neutron star (NS) and black-hole (BH) LMXBs \citep[see, e.g.,][for recent reviews]{done07,darias14}. 
Due to the lack of a simultaneous coverage with focusing X-ray instruments during the \inte\ revolution 1458 and the availability of 
spectra with sufficiently high statistics we could not investigate this possible spectral state transition in more details.   

The same spectral component that dominated all XRT spectra was also detected in the \xmm\ observation of the source 
carried out in 2015 March. The EPIC data clearly showed the presence of a number of absorption features, the most 
evident ones being centered at the known energies of the Fe XXV/Fe XXVI K$\alpha$ and K$\beta$ complexes. 
Similar features are observed in a number of LMXBs hosting either a NS or a BH, and are usually ascribed 
to the presence of an ionized absorbing material located above the accretion disk  
\citep{ueda98,kotani00,sidoli01,parmar02,boirin03,boirin04,boirin05,hyodo08}.   
Such absorbers are more likely to be observed in high inclination systems 
\citep[$\gtrsim$70~deg; see, e.g.,][]{ponti12,ponti14,dai14}, where our line of sight passes 
directly through the disk. In some cases, the absorption features show significant blueshifts, thus indicating that 
the ionized material is emerging from the system as a ``disk wind''  \citep[see, e.g.,][]{schulz02,kubota07,trigo07,iaria08}. 
Different driving mechanisms have been proposed to generate disk winds \citep[see, e.g.,][]{proga02}. 
These comprise a combination of radiation and thermal pressure, best suited to explain disk winds forming 
at distances of $\gtrsim$10$^{10}$-10$^{12}$~cm from the central compact object, or magnetic effects. The latter 
consider that the disk wind can be produced at relatively small distances from the compact object 
($\lesssim$10$^{9}$-10$^{10}$~cm) due to the pressure generated by the magnetic viscosity internal to
the disk or magnetocentrifugal forces \citep{miller06}. 
In all cases, the presence of an ionized absorbing material can be accounted 
for in the spectral analysis of high inclination LMXBs by using the {\sc warmabs} 
model in {\sc Xspec}. 
The fit with this model that we carried out on the \igr\ data in Sect.~\ref{sec:spectraph} 
revealed that the warm absorber in this source has a column density of $N_{\rm H}^{warmabs}\simeq10^{24}$~cm$^{-2}$, 
a turbulent velocity of $\sigma_v\simeq290$~km~s$^{-1}$, and a bulk velocity $v\simeq2200$~km~s$^{-1}$. 
The latter value indicates that this material is moving away from the disk and can thus be most likely 
associated to a disk wind. We note here that, even though the measured outflowing velocity in Table~\ref{tab:bestfit}  
is highly significant, the corresponding shift in energy of the involved lines 
is modest (about 0.7\% of the centroid energy) and is roughly of the same order of magnitude of the EPIC-pn calibration energy accuracy. 
For this reason, any interpretation about the presence of an outflowing velocity of the order of few thousands of km~s$^{-1}$ 
measured with the EPIC-pn has to be taken with caution.

From the estimated column density and ionization parameter of the warm absorber, we can 
also infer some information on the distance of this material from the compact object by using the definition of the ionization 
parameters $\xi=L_{\rm X}/n_e r^2$, where $L_{\rm X}$ is the source X-ray luminosity, 
$n_e$$\simeq$10$^{12}$~cm$^{-3}$ is the medium density assumed in the {\sc warmabs} model, and $r$ is the distance between 
the absorbing medium and the central source. Making use of the best fit values in Table~\ref{tab:bestfit}, we obtain 
$r\simeq2.0\times10^{10}$~cm, that is an intermediate value of the ionized absorber distance from the 
central compact object dividing thermal/radiative and magnetic models.   

We caution, however, that the distance to \igr\ is not known at present (thus potentially leading on large uncertainties 
on the estimated X-ray luminosity of the source) and that the above calculation depends on the density assumed 
for the partly ionized absorbing medium (see Sect.~\ref{sec:spectraph}). 
Disk winds are usually observed in bright LMXBs hosting a BH in the radio-free high soft state (HSS) or a NS with a luminosity 
$L_{\rm X}\gtrsim0.15L_{\rm Edd}$ \citep[here $L_{\rm Edd}$ is the Eddington luminosity for an accreting NS; see, e.g.,][and references 
therein]{ponti12,ponti14,ponti15}. The X-ray luminosity estimated from the unabsorbed \xmm\ flux of \igr\ is 
$L_{\rm X}\simeq4.3\times10^{36}(D/8~kpc)^2$~erg~s$^{-1}$, i.e about 2\% of the typical Eddington 
luminosity for a NS and a relatively low luminosity for a Galactic BH in the HSS \citep[see, e.g.][for a recent review]{done07}. 
This suggests that \igr\ might be located well beyond the Galactic Center, i.e. at distances $\gg$8~kpc. 
The lack of optical/radio/IR observations during the outburst of \igr\ (at the best of our knowledge) 
prevents, at present, a more accurate estimate of the source distance. We note that, as in order sources seen at high inclination, 
it is also possible that only a fraction of the intrinsic X-ray luminosity arrives to the observer \citep[see, e.g.,][]{burderi2010}, 
thus increasing our uncertainty concerning the source localization and the interpretation on its real nature. 
Concerning the assumed value of $n_e$, we commented in Sect.~\ref{sec:spectraph} that $\sim$10$^{12}$~cm$^{-3}$ is a reasonable estimate 
based on our knowledge of similar systems. Direct measurements of the plasma density in LMXBs 
displaying the presence of pronounced absorption features could be obtained so far only in two cases, and thus we cannot 
rule out substantially different values of $n_e$ \citep[see][and references therein]{trigo15}.  

At odds with other LMXBs seen at high inclinations and characterized by dense coronae \citep[see, e.g.][]{trigo13}, 
we found no evidence for either a broad iron emission line or a reflection component produced close to the inner disk boundary in \igr.\ 
The prominence of these features is a strong function of the system inclination and might become undetectable at $\gtrsim$80~deg 
\citep[see, e.g.,][]{garcia14}. We found evidence for an edge at $\sim$8.2~keV (see Table~\ref{tab:bestfit}) that could be due 
to Fe XXV, even though the measured energy is lower than expected \citep[8.828~keV; see, e.g.,][for a similar case]{disalvo09}. An alternative 
possibility is that the residuals around this energy are left due to a poorly accurate fit with the {\sc warmabs} model, especially 
in a energy range where the effective area of the EPIC-pn camera is decreasing rapidly and the signal-to-noise ratio 
of the data is far from being optimal (see Fig.~\ref{fig:bestfit}).   

The analysis of the EPIC-pn data revealed also the presence of two virtually rectangular eclipses in the source lightcurve, 
lasting about 820~s each. From the measurement of the two eclipse mid-epochs we provided an estimate of the source 
orbital period at 22620.5$^{+2.0}_{-1.8}$~s (1$\sigma$~c.l.; see Sect.~\ref{sec:eclipses}). 
The residual X-ray flux measured during the  
eclipses (about 6\% of the averaged uneclipsed flux) could in principle be explained by the presence of an 
accretion disk corona (ADC) located above the disk that scatters the X-rays from the central point-like source. However, 
as the X-ray spectrum extracted during the eclipse could be equally well fit with a {\sc diskbb} or a power-law model, we are 
unable to distinguish if this residual emission corresponds to a fraction of the scattered thermal 
X-rays from the disk or if there is also some significant reprocessing within the ADC with an eventual contribution 
from a scattering halo. This phenomenology is remarkably similar to that observed in the case of the NS LMXB EXO\,0748-676, 
which shows rectangular eclipses with a 4\% residual X-ray flux \citep[during X-ray active periods; see, e.g.,][]{parmar88}. 

The X-ray dips detected by \xmm\ from \igr\ also show remarkable 
similarities with the dips in EXO\,0748-676. In both systems, the dips precede in phase the X-ray eclipse and display 
complex structures varying from one orbit to the other. They also give rise to measurable changes in the hardness 
ratio of the source emission, which is interpreted in terms of increased absorption due to the presence of vertically 
extended structures in the disk at the impact point of the accretion stream from 
the donor star. It is worth noticing that EXO\,0748-676 also displayed clear evidence for the 
presence of an ionized absorbing material located above the accretion disk, even though no significant 
blueshift of the detected absorption features could be measured in that case \citep{ponti14}.  

Both the search for X-ray bursts and for pulsations resulted in non-detections, leaving the nature of the compact 
object in \igr\ unclear (see Sect.~\ref{sec:integral} and \ref{sec:timing}). 
As disk winds are nowadays understood to be an ubiquitous property of many NS and BH LMXBs observed at high inclination, 
this feature cannot be used to distinguish easily among the two possibilities in the case of \igr.\  
The analysis of the X-ray continuum from the source could, however, provide some 
additional clues. In particular, the normalization of the {\sc diskbb} model in the \xmm\ and \swift\ data is 
found to be always $\lesssim$20 and thus we can infer an approximate inner disk boundary of 12$(d/8~kpc)(\cos(\theta/85~deg))^{(-1/2)}$~km. 
Even when the usual color and torque-free boundary corrections are considered for the {\sc diskbb} component 
\citep[see, e.g.,][and references therein]{gierl99,nowak12}, 
the inner disk boundary estimated from the X-ray spectrum of \igr\ remains uncomfortably small for a black-hole 
(less than 2~$r_{\rm g}$ for a 3~$M_{\odot}$ BH). Although this does not allow us to 
draw a firm conclusion about the nature of the compact object in \igr,\ we conclude that the 
NS hypothesis is favoured (see also Zdziarski et al. 2016, in preparation).

\section*{Acknowledgements}
 
We thank N. Schartel and the \xmm\ team for having promptly performed the ToO observation analyzed in this paper.  
We are indebted to the Swift PI and operations team for the continuous support during the monitoring campaign 
of X-ray transients. AAZ and PP  
have been supported in part by the Polish NCN grants 2012/04/M/ST9/00780 and 2013/10/M/ST9/00729. 
AP is supported by a Juan de la Cierva fellowship, and acknowledges grants AYA2012-39303, SGR2009-811, and iLINK2011-0303. 
PR acknowledges contract ASI-INAF I/004/11/0. GP acknowledges support by the Bundesministerium f\"ur Wirtschaft und 
Technologie/Deutsches Zentrum f\"ur Luft- und Raumfahrt (BMWI/DLR, FKZ 50 OR 1408)
and the Max Planck Society. LB e TD acknowledge contract ASI-INAF. 

\bibliographystyle{aa}
\bibliography{J17451}

\begin{thebibliography}{54}
\expandafter\ifx\csname natexlab\endcsname\relax\def\natexlab#1{#1}\fi

\bibitem[{{Altamirano} {et~al.}(2014){Altamirano}, {Wijnands}, {Heinke}, \&
  {Bahramian}}]{Altamirano14}
{Altamirano}, D., {Wijnands}, R., {Heinke}, C.~O., \& {Bahramian}, A. 2014, The
  Astronomer's Telegram, 6469, 1

\bibitem[{{Bahramian} {et~al.}(2015{\natexlab{a}}){Bahramian}, {Heinke},
  {Altamirano}, \& {Wijnands}}]{Bahramian15a}
{Bahramian}, A., {Heinke}, C.~O., {Altamirano}, D., \& {Wijnands}, R.
  2015{\natexlab{a}}, The Astronomer's Telegram, 7570, 1

\bibitem[{{Bahramian} {et~al.}(2014{\natexlab{a}}){Bahramian}, {Heinke},
  {Beardmore}, {Altamirano}, \& {Wijnands}}]{Bahramian15c}
{Bahramian}, A., {Heinke}, C.~O., {Beardmore}, A.~P., {Altamirano}, D., \&
  {Wijnands}, R. 2014{\natexlab{a}}, The Astronomer's Telegram, 6501, 1

\bibitem[{{Bahramian} {et~al.}(2014{\natexlab{b}}){Bahramian}, {Heinke},
  {Wijnands}, \& {Altamirano}}]{Bahramian15d}
{Bahramian}, A., {Heinke}, C.~O., {Wijnands}, R., \& {Altamirano}, D.
  2014{\natexlab{b}}, The Astronomer's Telegram, 6486, 1

\bibitem[{{Bahramian} {et~al.}(2015{\natexlab{b}}){Bahramian}, {Heinke},
  {Wijnands}, \& {Altamirano}}]{Bahramian15b}
{Bahramian}, A., {Heinke}, C.~O., {Wijnands}, R., \& {Altamirano}, D.
  2015{\natexlab{b}}, The Astronomer's Telegram, 7028, 1

\bibitem[{{Boirin} {et~al.}(2005){Boirin}, {M{\'e}ndez}, {D{\'{\i}}az Trigo},
  {Parmar}, \& {Kaastra}}]{boirin05}
{Boirin}, L., {M{\'e}ndez}, M., {D{\'{\i}}az Trigo}, M., {Parmar}, A.~N., \&
  {Kaastra}, J.~S. 2005, \aap, 436, 195

\bibitem[{{Boirin} \& {Parmar}(2003)}]{boirin03}
{Boirin}, L. \& {Parmar}, A.~N. 2003, \aap, 407, 1079

\bibitem[{{Boirin} {et~al.}(2004){Boirin}, {Parmar}, {Barret}, {Paltani}, \&
  {Grindlay}}]{boirin04}
{Boirin}, L., {Parmar}, A.~N., {Barret}, D., {Paltani}, S., \& {Grindlay},
  J.~E. 2004, \aap, 418, 1061

\bibitem[{{Bozzo} {et~al.}(2007){Bozzo}, {Falanga}, {Papitto}, {Stella},
  {Perna}, {Lazzati}, {Israel}, {Campana}, {Mangano}, {di Salvo}, \&
  {Burderi}}]{bozzo07}
{Bozzo}, E., {Falanga}, M., {Papitto}, A., {et~al.} 2007, \aap, 476, 301

\bibitem[{{Burderi} {et~al.}(2010){Burderi}, {Di Salvo}, {Riggio}, {Papitto},
  {Iaria}, {D'A{\`i}}, \& {Menna}}]{burderi2010}
{Burderi}, L., {Di Salvo}, T., {Riggio}, A., {et~al.} 2010, \aap, 515, A44

\bibitem[{{Burrows} {et~al.}(2005){Burrows}, {Hill}, {Nousek}, {Kennea},
  {Wells}, {Osborne}, {Abbey}, {Beardmore}, {Mukerjee}, {Short}, {Chincarini},
  {Campana}, {Citterio}, {Moretti}, {Pagani}, {Tagliaferri}, {Giommi},
  {Capalbi}, {Tamburelli}, {Angelini}, {Cusumano}, {Br{\"a}uninger}, {Burkert},
  \& {Hartner}}]{burrows05}
{Burrows}, D.~N., {Hill}, J.~E., {Nousek}, J.~A., {et~al.} 2005, \ssr, 120, 165

\bibitem[{{Chakrabarty} {et~al.}(2014){Chakrabarty}, {Jonker}, \&
  {Markwardt}}]{Chakrabarty15}
{Chakrabarty}, D., {Jonker}, P.~G., \& {Markwardt}, C.~B. 2014, The
  Astronomer's Telegram, 6533, 1

\bibitem[{{Chenevez} {et~al.}(2014){Chenevez}, {Vandbaek Kroer},
  {Budtz-Jorgensen}, {Brandt}, {Lund}, {Westergaard}, {Kuulkers}, \&
  {Wilms}}]{Chenevez14}
{Chenevez}, J., {Vandbaek Kroer}, L., {Budtz-Jorgensen}, C., {et~al.} 2014, The
  Astronomer's Telegram, 6451, 1

\bibitem[{{Courvoisier} {et~al.}(2003){Courvoisier}, {Walter}, {Beckmann},
  {Dean}, {Dubath}, {Hudec}, {Kretschmar}, {Mereghetti}, {Montmerle},
  {Mowlavi}, {Paltani}, {Preite Martinez}, {Produit}, {Staubert}, {Strong},
  {Swings}, {Westergaard}, {White}, {Winkler}, \& {Zdziarski}}]{courvoisier03}
{Courvoisier}, T., {Walter}, R., {Beckmann}, V., {et~al.} 2003, \aap, 411, L53

\bibitem[{{D'A{\`i}} {et~al.}(2015){D'A{\`i}}, {Di Salvo}, {Iaria},
  {Garc{\'{\i}}a}, {Sanna}, {Pintore}, {Riggio}, {Burderi}, {Bozzo}, {Dauser},
  {Matranga}, {Galiano}, \& {Robba}}]{dai15}
{D'A{\`i}}, A., {Di Salvo}, T., {Iaria}, R., {et~al.} 2015, \mnras, 449, 4288

\bibitem[{{D'A{\`i}} {et~al.}(2014){D'A{\`i}}, {Iaria}, {Di Salvo}, {Riggio},
  {Burderi}, \& {Robba}}]{dai14}
{D'A{\`i}}, A., {Iaria}, R., {Di Salvo}, T., {et~al.} 2014, \aap, 564, A62

\bibitem[{{di Salvo} {et~al.}(2009){di Salvo}, {D'A{\'{\i}}}, {Iaria},
  {Burderi}, {Dov{\v c}iak}, {Karas}, {Matt}, {Papitto}, {Piraino}, {Riggio},
  {Robba}, \& {Santangelo}}]{disalvo09}
{di Salvo}, T., {D'A{\'{\i}}}, A., {Iaria}, R., {et~al.} 2009, \mnras, 398,
  2022

\bibitem[{{D{\'{\i}}az Trigo} \& {Boirin}(2013)}]{trigo13}
{D{\'{\i}}az Trigo}, M. \& {Boirin}, L. 2013, Acta Polytechnica, 53, 659

\bibitem[{{D{\'{\i}}az Trigo} \& {Boirin}(2015)}]{trigo15}
{D{\'{\i}}az Trigo}, M. \& {Boirin}, L. 2015, ArXiv e-prints

\bibitem[{{D{\'{\i}}az Trigo} {et~al.}(2014){D{\'{\i}}az Trigo}, {Migliari},
  {Miller-Jones}, \& {Guainazzi}}]{trigo14}
{D{\'{\i}}az Trigo}, M., {Migliari}, S., {Miller-Jones}, J.~C.~A., \&
  {Guainazzi}, M. 2014, \aap, 571, A76

\bibitem[{{D{\'{\i}}az Trigo} {et~al.}(2007){D{\'{\i}}az Trigo}, {Parmar},
  {Miller}, {Kuulkers}, \& {Caballero-Garc{\'{\i}}a}}]{trigo07}
{D{\'{\i}}az Trigo}, M., {Parmar}, A.~N., {Miller}, J., {Kuulkers}, E., \&
  {Caballero-Garc{\'{\i}}a}, M.~D. 2007, \aap, 462, 657

\bibitem[{{Done} {et~al.}(2007){Done}, {Gierli{\'n}ski}, \& {Kubota}}]{done07}
{Done}, C., {Gierli{\'n}ski}, M., \& {Kubota}, A. 2007, \aapr, 15, 1

\bibitem[{{Garc{\'{\i}}a} {et~al.}(2014){Garc{\'{\i}}a}, {Dauser}, {Lohfink},
  {Kallman}, {Steiner}, {McClintock}, {Brenneman}, {Wilms}, {Eikmann},
  {Reynolds}, \& {Tombesi}}]{garcia14}
{Garc{\'{\i}}a}, J., {Dauser}, T., {Lohfink}, A., {et~al.} 2014, \apj, 782, 76

\bibitem[{{Gierli{\'n}ski} {et~al.}(1999){Gierli{\'n}ski}, {Zdziarski},
  {Poutanen}, {Coppi}, {Ebisawa}, \& {Johnson}}]{gierl99}
{Gierli{\'n}ski}, M., {Zdziarski}, A.~A., {Poutanen}, J., {et~al.} 1999,
  \mnras, 309, 496

\bibitem[{{Heinke} {et~al.}(2014){Heinke}, {Bahramian}, {Altamirano}, \&
  {Wijnands}}]{Heinke14}
{Heinke}, C.~O., {Bahramian}, A., {Altamirano}, D., \& {Wijnands}, R. 2014, The
  Astronomer's Telegram, 6459, 1

\bibitem[{{Hyodo} {et~al.}(2008){Hyodo}, {Ueda}, {Yuasa}, {Maeda}, {Makishima},
  \& {Koyama}}]{hyodo08}
{Hyodo}, Y., {Ueda}, Y., {Yuasa}, T., {et~al.} 2008, ArXiv e-prints

\bibitem[{{Iaria} {et~al.}(2008){Iaria}, {D'A{\'{\i}}}, {Lavagetto}, {Di
  Salvo}, {Robba}, \& {Burderi}}]{iaria08}
{Iaria}, R., {D'A{\'{\i}}}, A., {Lavagetto}, G., {et~al.} 2008, \apj, 673, 1033

\bibitem[{{Jaisawal} {et~al.}(2015){Jaisawal}, {Homan}, {Naik}, \&
  {Jonker}}]{Jaisawal15}
{Jaisawal}, G.~K., {Homan}, J., {Naik}, S., \& {Jonker}, P. 2015, The
  Astronomer's Telegram, 7361, 1

\bibitem[{{Jansen} {et~al.}(2001){Jansen}, {Lumb}, {Altieri}, {Clavel}, {Ehle},
  {Erd}, {Gabriel}, {Guainazzi}, {Gondoin}, {Much}, {Munoz}, {Santos},
  {Schartel}, {Texier}, \& {Vacanti}}]{jansen01}
{Jansen}, F., {Lumb}, D., {Altieri}, B., {et~al.} 2001, \aap, 365, L1

\bibitem[{{Johnston} \& {Kulkarni}(1991)}]{johnston11}
{Johnston}, H.~M. \& {Kulkarni}, S.~R. 1991, \apj, 368, 504

\bibitem[{{Kotani} {et~al.}(2000){Kotani}, {Ebisawa}, {Dotani}, {Inoue},
  {Nagase}, {Tanaka}, \& {Ueda}}]{kotani00}
{Kotani}, T., {Ebisawa}, K., {Dotani}, T., {et~al.} 2000, \apj, 539, 413

\bibitem[{{Kubota} {et~al.}(2007){Kubota}, {Dotani}, {Cottam}, {Kotani},
  {Done}, {Ueda}, {Fabian}, {Yasuda}, {Takahashi}, {Fukazawa}, {Yamaoka},
  {Makishima}, {Yamada}, {Kohmura}, \& {Angelini}}]{kubota07}
{Kubota}, A., {Dotani}, T., {Cottam}, J., {et~al.} 2007, \pasj, 59, 185

\bibitem[{{Lebrun} {et~al.}(2003){Lebrun}, {Leray}, {Lavocat}, {Cr{\'e}tolle},
  {Arqu{\`e}s}, {Blondel}, {Bonnin}, {Bou{\`e}re}, {Cara}, {Chaleil}, {Daly},
  {Desages}, {Dzitko}, {Horeau}, {Laurent}, {Limousin}, {Mathy}, {Mauguen},
  {Meignier}, {Molini{\'e}}, {Poindron}, {Rouger}, {Sauvageon}, \&
  {Tourrette}}]{lebrun03}
{Lebrun}, F., {Leray}, J.~P., {Lavocat}, P., {et~al.} 2003, \aap, 411, L141

\bibitem[{{Lund} {et~al.}(2003){Lund}, {Budtz-J{\o}rgensen}, {Westergaard},
  {Brandt}, {Rasmussen}, {Hornstrup}, {Oxborrow}, {Chenevez}, {Jensen},
  {Laursen}, {Andersen}, {Mogensen}, {Rasmussen}, {Om{\o}}, {Pedersen},
  {Polny}, {Andersson}, {Andersson}, {K{\"a}m{\"a}r{\"a}inen}, {Vilhu},
  {Huovelin}, {Maisala}, {Morawski}, {Juchnikowski}, {Costa}, {Feroci},
  {Rubini}, {Rapisarda}, {Morelli}, {Carassiti}, {Frontera}, {Pelliciari},
  {Loffredo}, {Mart{\'{\i}}nez N{\'u}{\~n}ez}, {Reglero}, {Velasco}, {Larsson},
  {Svensson}, {Zdziarski}, {Castro-Tirado}, {Attina}, {Goria}, {Giulianelli},
  {Cordero}, {Rezazad}, {Schmidt}, {Carli}, {Gomez}, {Jensen}, {Sarri},
  {Tiemon}, {Orr}, {Much}, {Kretschmar}, \& {Schnopper}}]{lund03}
{Lund}, N., {Budtz-J{\o}rgensen}, C., {Westergaard}, N.~J., {et~al.} 2003,
  \aap, 411, L231

\bibitem[{{Miller} {et~al.}(2006){Miller}, {Raymond}, {Fabian}, {Steeghs},
  {Homan}, {Reynolds}, {van der Klis}, \& {Wijnands}}]{miller06}
{Miller}, J.~M., {Raymond}, J., {Fabian}, A., {et~al.} 2006, \nat, 441, 953

\bibitem[{{Mu{\~n}oz-Darias} {et~al.}(2014){Mu{\~n}oz-Darias}, {Fender},
  {Motta}, \& {Belloni}}]{darias14}
{Mu{\~n}oz-Darias}, T., {Fender}, R.~P., {Motta}, S.~E., \& {Belloni}, T.~M.
  2014, \mnras, 443, 3270

\bibitem[{{Nowak} {et~al.}(2012){Nowak}, {Wilms}, {Pottschmidt}, {Schulz},
  {Maitra}, \& {Miller}}]{nowak12}
{Nowak}, M.~A., {Wilms}, J., {Pottschmidt}, K., {et~al.} 2012, \apj, 744, 107

\bibitem[{{Parmar} {et~al.}(2002){Parmar}, {Oosterbroek}, {Boirin}, \&
  {Lumb}}]{parmar02}
{Parmar}, A.~N., {Oosterbroek}, T., {Boirin}, L., \& {Lumb}, D. 2002, \aap,
  386, 910

\bibitem[{{Parmar} \& {White}(1988)}]{parmar88}
{Parmar}, A.~N. \& {White}, N.~E. 1988, \memsai, 59, 147

\bibitem[{{Pintore} {et~al.}(2015){Pintore}, {Di Salvo}, {Bozzo}, {Sanna},
  {Burderi}, {D'A{\`i}}, {Riggio}, {Scarano}, \& {Iaria}}]{pintore15}
{Pintore}, F., {Di Salvo}, T., {Bozzo}, E., {et~al.} 2015, \mnras, 450, 2016

\bibitem[{{Pintore} {et~al.}(2014){Pintore}, {Sanna}, {Di Salvo}, {Guainazzi},
  {D'A{\`i}}, {Riggio}, {Burderi}, {Iaria}, \& {Robba}}]{pintore14}
{Pintore}, F., {Sanna}, A., {Di Salvo}, T., {et~al.} 2014, \mnras, 445, 3745

\bibitem[{{Piraino} {et~al.}(2012){Piraino}, {Santangelo}, {Kaaret},
  {M{\"u}ck}, {D'A{\`i}}, {Di Salvo}, {Iaria}, {Robba}, {Burderi}, \&
  {Egron}}]{piraino12}
{Piraino}, S., {Santangelo}, A., {Kaaret}, P., {et~al.} 2012, \aap, 542, L27

\bibitem[{{Ponti} {et~al.}(2015){Ponti}, {Bianchi}, {Mu{\~n}oz-Darias}, {De
  Marco}, {Dwelly}, {Fender}, {Nandra}, {Rea}, {Mori}, {Haggard}, {Heinke},
  {Degenaar}, {Aramaki}, {Clavel}, {Goldwurm}, {Hailey}, {Israel}, {Morris},
  {Rushton}, \& {Terrier}}]{ponti15}
{Ponti}, G., {Bianchi}, S., {Mu{\~n}oz-Darias}, T., {et~al.} 2015, \mnras, 446,
  1536

\bibitem[{{Ponti} {et~al.}(2012){Ponti}, {Fender}, {Begelman}, {Dunn},
  {Neilsen}, \& {Coriat}}]{ponti12}
{Ponti}, G., {Fender}, R.~P., {Begelman}, M.~C., {et~al.} 2012, \mnras, 422,
  L11

\bibitem[{{Ponti} {et~al.}(2014){Ponti}, {Mu{\~n}oz-Darias}, \&
  {Fender}}]{ponti14}
{Ponti}, G., {Mu{\~n}oz-Darias}, T., \& {Fender}, R.~P. 2014, \mnras, 444, 1829

\bibitem[{{Proga} \& {Kallman}(2002)}]{proga02}
{Proga}, D. \& {Kallman}, T.~R. 2002, \apj, 565, 455

\bibitem[{{Romano} {et~al.}(2006){Romano}, {Campana}, {Chincarini}, {Cummings},
  {Cusumano}, {Holland}, {Mangano}, {Mineo}, {Page}, {Pal'Shin}, {Rol},
  {Sakamoto}, {Zhang}, {Aptekar}, {Barbier}, {Barthelmy}, {Beardmore}, {Boyd},
  {Burrows}, {Capalbi}, {Fenimore}, {Frederiks}, {Gehrels}, {Giommi}, {Goad},
  {Godet}, {Golenetskii}, {Guetta}, {Kennea}, {La Parola}, {Malesani},
  {Marshall}, {Moretti}, {Nousek}, {O'Brien}, {Osborne}, {Perri}, \&
  {Tagliaferri}}]{Romano2006:060124}
{Romano}, P., {Campana}, S., {Chincarini}, G., {et~al.} 2006, \aap, 456, 917

\bibitem[{{Schulz} \& {Brandt}(2002)}]{schulz02}
{Schulz}, N.~S. \& {Brandt}, W.~N. 2002, \apj, 572, 971

\bibitem[{{Sidoli} {et~al.}(2001){Sidoli}, {Oosterbroek}, {Parmar}, {Lumb}, \&
  {Erd}}]{sidoli01}
{Sidoli}, L., {Oosterbroek}, T., {Parmar}, A.~N., {Lumb}, D., \& {Erd}, C.
  2001, \aap, 379, 540

\bibitem[{{Ubertini} {et~al.}(2003){Ubertini}, {Lebrun}, {Di Cocco}, {Bazzano},
  {Bird}, {Broenstad}, {Goldwurm}, {La Rosa}, {Labanti}, {Laurent}, {Mirabel},
  {Quadrini}, {Ramsey}, {Reglero}, {Sabau}, {Sacco}, {Staubert}, {Vigroux},
  {Weisskopf}, \& {Zdziarski}}]{ubertini03}
{Ubertini}, P., {Lebrun}, F., {Di Cocco}, G., {et~al.} 2003, \aap, 411, L131

\bibitem[{{Ueda} {et~al.}(1998){Ueda}, {Inoue}, {Tanaka}, {Ebisawa}, {Nagase},
  {Kotani}, \& {Gehrels}}]{ueda98}
{Ueda}, Y., {Inoue}, H., {Tanaka}, Y., {et~al.} 1998, \apj, 492, 782

\bibitem[{{Vaughan} {et~al.}(1994){Vaughan}, {van der Klis}, {Wood}, {Norris},
  {Hertz}, {Michelson}, {van Paradijs}, {Lewin}, {Mitsuda}, \&
  {Penninx}}]{vaughan94}
{Vaughan}, B.~A., {van der Klis}, M., {Wood}, K.~S., {et~al.} 1994, \apj, 435,
  362

\bibitem[{{Vaughan} {et~al.}(2006){Vaughan}, {Goad}, {Beardmore}, {O'Brien},
  {Osborne}, {Page}, {Barthelmy}, {Burrows}, {Campana}, {Cannizzo}, {Capalbi},
  {Chincarini}, {Cummings}, {Cusumano}, {Giommi}, {Godet}, {Hill}, {Kobayashi},
  {Kumar}, {La Parola}, {Levan}, {Mangano}, {M{\'e}sz{\'a}ros}, {Moretti},
  {Morris}, {Nousek}, {Pagani}, {Palmer}, {Racusin}, {Romano}, {Tagliaferri},
  {Zhang}, \& {Gehrels}}]{vaughan2006:050315}
{Vaughan}, S., {Goad}, M.~R., {Beardmore}, A.~P., {et~al.} 2006, \apj, 638, 920

\bibitem[{{Wood} {et~al.}(1991){Wood}, {Norris}, {Hertz}, {Vaughan},
  {Michelson}, {Mitsuda}, {Lewin}, {van Paradijs}, {Penninx}, \& {van der
  Klis}}]{wood91}
{Wood}, K.~S., {Norris}, J.~P., {Hertz}, P., {et~al.} 1991, \apj, 379, 295

\end{thebibliography}

\end{document}